\documentclass[namedreferences,hyperref,optionalrh,solaromanenum]{spr-sola}
\usepackage{graphicx}                    % For eps figures, newer & more powerfull
\usepackage{epstopdf}
\usepackage{color}                       % For color text: \color command
\usepackage{tabularx}
\usepackage{rotating}

\chardef\us=`\_

\begin{document}

\begin{frontmatter}
\title{Latitudinal dependence of variations in the frequencies of solar oscillations above the acoustic cut-off}

\author[addressref={aff1},corref,email={laura.millson@warwick.ac.uk}]{\inits{L. J.}\fnm{Laura Jade}~\snm{Millson}\orcid{0009-0003-4254-2676}}
\author[addressref=aff1,email={a-m.broomhall@warwick.ac.uk}]{\inits{A-M.}\fnm{Anne-Marie}~\snm{Broomhall}\orcid{0000-0002-5209-9378}}
\author[addressref=aff1]{\inits{T.}\fnm{Tishtrya}~\snm{Mehta}\orcid{0000-0002-4875-9142}}
%\author{\inits{}\fnm{}~\lnm{}\orcid{}}
%   NOTE:  Just one corresponding author [corref]
\address[id=aff1]{Centre for Fusion, Space and Astrophysics (CFSA), Department of Physics, University of Warwick, Coventry, CV4 7AL, UK}

\runningauthor{Millson et al.}
\runningtitle{Pseudo-mode frequencies across solar latitude}

\begin{abstract}
At high frequencies beyond the acoustic cut-off, a peak-like structure is visible in the solar power spectrum. Known as the pseudo-modes, their frequencies have been shown to vary in anti-phase with solar magnetic activity. In this work, we determined temporal variations in these frequencies across the solar disc, with the aim of identifying any potential latitudinal dependence of pseudo-mode frequency shifts. We utilised nearly 22 years of spatially resolved GONG data for all azimuthal orders, $\textit{m}$, for harmonic degrees 0 $\leq$ $\textit{l}$ $\leq$ 200, and determined shifts using the resampled periodogram method. Periodogram realisations were created from overlapping, successive 216d-long segments in time, and cropped to 5600-6800$\mu$Hz. Cross-correlation functions were then repeatedly generated between these realisations to identify any variation in frequency and the uncertainty. We categorised each mode by its latitudinal sensitivity and used this categorisation to produce average frequency shifts for different latitude bands (15$^\circ$ and 5$^\circ$ in size) which were compared to magnetic proxies, the $F_{\mathrm{10.7}}$ index and GONG synoptic maps. Morphological differences in the pseudo-mode shifts between different latitudes were found, which were most pronounced during the rise to solar maximum where shifts reach their minimum values. At all latitudes, shift behaviour was strongly in anti-correlation with the activity proxy. Additionally, periodicities shorter than the 11-year cycle were observed. Wavelet analysis was used to identify a periodicity of four years at all latitudes.
\end{abstract}
\end{frontmatter}
%-------------------------------------------------

\section{Introduction}
     \label{Introduction} 

The Sun exhibits periodic magnetic activity of varying lengths and amplitudes. One such phenomenon, the 22-year solar cycle, features two 11-year sunspot cycles where the polarity is reversed in successive cycles. This periodic variation is reflected in solar proxies and through helioseismic studies. Acoustic p-mode frequencies have been found to vary in-phase with the 11-year solar cycle (frequencies shifted to higher values as magnetic activity increased) (\citealt{woodard1985change}; \citealt{palle1989solar}; \citealt{libbrecht1990solar}; \citealt{elsworth1990variation}; \citealt{jimenez1998solar}; see \citealt{broomhall2014sun} and \citealt{basu2016global} and references therein for a full review), and the p mode heights in anti-phase (\citealt{palle1990progress}; \citealt{chaplin2000variations}; \citealt{2018SoPh..293..151K}; \citealt{2019ApJ...877..148K}). The amplitude of frequency shifts have also been shown to be greater for higher frequency modes (\citealt{libbrecht1990solar}; \citealt{anguera1992low}; \citealt{2017SoPh..292...67B}). As higher frequency modes have higher upper turning points in the Sun, frequency variations with the magnetic activity cycle are thought to result from changes to the acoustic properties of the uppermost layer of the solar interior, just beneath the photosphere (\citealt{libbrecht1990solar}; \citealt{goldreich1991implications}; \citealt{nishizawa1995implication}; \citealt{2012ApJ...758...43B}; \citealt{2018MNRAS.480L..79H}).

Shorter-term modulation of the solar magnetic activity cycle has also been identified in numerous solar activity proxies and through seismology. Known as quasi-biennial oscillations (QBO), these oscillations have periods on timescales of 0.6-4 years (\citealt{bazilevskaya2014combined}; \citealt{mehta2022cycle}; \citealt{jain2023helioseismic}). The seismic QBO has been observed in p-mode frequency shifts with an $\approx$2-year periodicity (\citealt{broomhall2009current}; \citealt{fletcher2010seismic}; \citealt{broomhall2011short}), and is modulated by the 11-year solar cycle (it has a greater amplitude at solar maximum, but is still visible when activity is at a minimum) \citep{broomhall2011short}. There is also evidence to suggest that the frequency dependence of this seismic QBO is weaker than the dependence of the 11-year solar cycle (\citealt{broomhall2012quasi}; \citealt{simoniello2012seismic}), implying perturbations generating the seismic QBO are deeper in the solar interior compared to perturbations of the uppermost layer for the 11-year cycle.

Throughout the solar cycle, sunspots drift from mid-latitudes ($\pm$30$^\circ$) to lower latitudes ($\pm$5-10$^\circ$). The mapping of sunspot locations over magnetic cycles results in the well-known butterfly diagram. This pattern is also visible when mapping the magnitude of p-mode frequency shifts over solar latitude and time (\citealt{howe2002localizing}; \citealt{broomhall2014sun}). To investigate whether this latitudinal progression of the solar cycle is observed in the acoustic p modes, \citet{simoniello2016new} searched for temporal variations in p-mode frequencies as a function of latitude throughout Solar Cycle 23 and the rising phase to Cycle 24 maximum (June 1995 - July 2013) utilising intermediate (20 $\leq$ \textit{l} $\leq$ 147) and high (180 $\leq$ \textit{l} $\leq$ 1000) harmonic degree modes. The authors identified that whilst frequency shifts for all latitudes above $\pm$15$^\circ$ increased with the rise in magnetic activity, there was a delayed onset (1-2 years) in the upturn of frequency shift values for the lowest latitude band $\pm$0$^\circ$-15$^\circ$. Frequency shifts in the $\pm$0$^\circ$-15$^\circ$ band were found to have the fastest rise time (to solar maximum), and a delay in the decrease of the frequency shift amplitude after solar maximum (compared to the other latitudes). This delay resulted in an overlap of successive cycles at the frequency shift minimum. Above $\pm$15$^\circ$, the authors identified a double peak structure at solar maximum (with peaks in 2000 and 2002), which was interpreted as a manifestation of the QBO.

In this work, we aim to conduct the first study on the latitudinal behaviour of frequency shifts of the pseudo-modes. In a solar power spectrum, we observe a mode-like pattern which extends beyond the acoustic cut-off frequency, $\nu_{\mathrm{ac}}$ (\citealt{jefferies1988helioseismology}; \citealt{libbrecht1988solar}; \citealt{kumar1990observed}; \citealt{duvall1991measurements}). This structure, known as the pseudo-modes, is a result of waves travelling directly towards the observer, and indirectly upon refraction in the solar interior or at the back of the star, which interfere 'geometrically' and are observed projected onto spherical harmonics (\citealt{kumar1990observed}; \citealt{kumar1991location}; \citealt{vorontsov1998acoustic}; \citealt{garcia1998high}).

Pseudo-modes are an interesting tool for studies on solar magnetic activity. Firstly, for the Sun, pseudo-mode frequency variations are strongly anti-correlated with solar magnetic activity. The frequencies of the pseudo-modes have been shown to vary in opposition with the 11-year magnetic cycle between solar minimum and maximum (\citealt{ronan1994solar}; \citealt{rhodes2011temporal}), and as a function of time \citep{kosak2022multi}. In \citet{kosak2022multi}, the authors determined pseudo-mode frequency shifts between time segments of 100 days (overlapping by 50 days) over nearly 20 years of GONG data averaged over all azimuthal orders. The pseudo-mode frequencies were found to vary significantly in anti-phase with the solar cycle for harmonic degrees 4~$\leq$~\textit{l}~$\leq$~200. Secondly, the amplitude of the shift of the pseudo-mode frequencies is larger than that of the p-mode frequencies. \citet{kosak2022multi} showed pseudo-mode frequency shift amplitude to vary by $\approx$1.5$\mu$Hz throughout Solar Cycles 23 and 24.

Currently, we only have a global view of the behaviour of pseudo-mode frequency shifts, where shifts are known to be anti-correlated with the solar magnetic activity cycle. However, prior analysis of the latitudinal dependence of p-mode frequency shifts has shown differences in the progression of the solar cycle below $\pm$15$^\circ$ between shifts and sunspots, providing insight on magnetic field structures and solar dynamo mechanisms \citep{simoniello2016new}. Latitudinal analysis of pseudo-mode frequency shifts may, therefore, better constrain the origin and nature of variations in pseudo-mode frequencies, and the temporal interplay between shifts and magnetic activity. In this study, we utilised 22 years of GONG data to determine the behaviour of the solar pseudo-mode frequency shifts over time as a function of latitude across the solar disc. We present our methods in Section \ref{Methodology}, the results in Section \ref{Results}, and our conclusions in Section \ref{Conclusion}.

\begin{figure} 
    \centerline{\includegraphics[width=1\textwidth, clip=, draft=false]{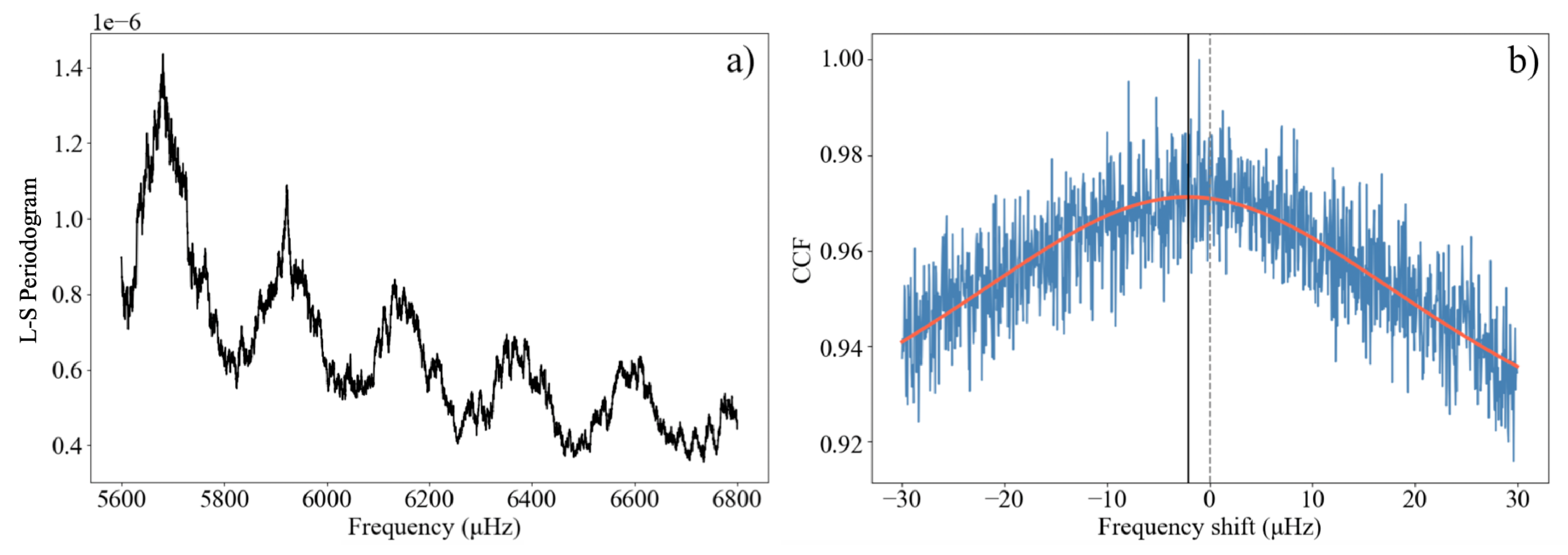}}
    \caption{Resampled periodogram cross-correlation (CCF) approach used to determine frequency shifts of the pseudo-modes between successive time segments. (a) Smoothed (10$\mu$Hz) Lomb-Scargle periodogram showing the peak-like structure of the pseudo-modes (located at frequencies beyond $\nu_{\mathrm{ac}}$, 5600-6800$\mu$Hz shown here) from resolved GONG solar data for mode $\textit{l}$=150, $\textit{m}$=0, and (b) the CCF generated between a reference and a comparison periodogram realisation using the resampling approach. The CCF was cropped to $\pm$30$\mu$Hz, a Lorentzian was fitted (red line), and the frequency shift was recorded as the centre of the Lorentzian (black line, dashed grey line shows 0$\mu$Hz).}
    \label{Figure 1}
\end{figure}

\section{Methodology}
     \label{Methodology} 

\subsection{Data preparation}
    \label{Data preparation} 

In this work, we searched for temporal variations in the frequency shifts of the pseudo-modes at different latitudes across the solar disc. For this, we utilised data from the Global Oscillations Network Group (GONG, \citealt{harvey1996global}, \url{https://gong.nso.edu}).

GONG consists of a network of six sites around the Earth which continuously monitor the Sun. Network merged timeseries (mrvmt) are produced by merging individual site timeseries, and concatenating 36 days of observations (known as a GONG month) \citep{hill1996solar}. We downloaded 223 of these timeseries, covering the time between 16th June 2001 to 7th June 2023. We started our analysis from 16th June 2001 onwards, as this is when the cameras were upgraded from 256~x~256~pixels to 1024~x~1024~pixels (with 60s cadence).

The 36-day timeseries is spatially resolved for harmonic degrees 0~$\leq$~\textit{l}~$\leq$~200. For each \textit{l}, there are 2\textit{l}+1 azimuthal orders, \textit{m}, such that for \textit{l}=200, there are 401 azimuthal orders ranging from -200~$\leq$~\textit{m}~$\leq$~200.

\subsection{Temporal frequency shifts}
\label{sec:freqshiftsmethod} 

To search for temporal changes in frequency shifts of the pseudo-modes, and the uncertainty, we made use of \citet{kiefer2017}'s resampled periodogram approach. 
For each azimuthal order for every harmonic degree over 0~$\leq$~\textit{l}~$\leq$~200, we concatenate six 36-day timeseries to create segments 216 days in length. Each successive segment overlaps the prior by half (108 days). We considered the first 216-day long segment (16th June 2001-17th January 2002) the reference segment, and so all frequency shifts are relative to this. A Lomb-Scargle (L-S) periodogram was then generated for both the reference segment and the comparison segment. Both periodograms were smoothed using a boxcar window with a width of 10$\mu$Hz, following the approach of \citet{kosak2022multi}. Each periodogram was then cropped to the range of frequencies in which we expected to observe the signature of the pseudo-modes. For this analysis, the range 5600-6800$\mu$Hz was used. The frequency 5600$\mu$Hz was chosen as a lower limit to ensure no p modes were included in the analysis ($\nu_{\mathrm{ac,\odot}}$$\approx$5100$\mu$Hz, \citealt{jimenez2006estimation}). The upper limit, 6800$\mu$Hz, was selected so as many pseudo-mode peaks (6-8 peaks separated by $\approx$$\Delta\nu$) would be included in the analysis as possible, but that the upper limit was below both the frequency at which frequency shifts have been shown to switch to be in-phase with magnetic activity ($\approx$6800$\mu$Hz) \citep{rhodes2011temporal} and the Nyquist frequency ($\approx$8333$\mu$Hz).
An example of a L-S periodogram for the solar pseudo-mode region for mode \textit{l}=150, \textit{m}=0 is shown in Figure \ref{Figure 1}(a). To generate the periodogram realisations, we then took the square root of the periodograms and multiplied by a zero mean normal distribution, \textit{N}(0, 1). This was repeated for both a 'real' and 'imaginary' instance. The modulus of the complex value was then determined, and this absolute value was squared. This generated new periodogram realisations for each segment, which retained the statistical properties ($\chi^2$ with 2dof) of the original periodograms. A cross-correlation function (CCF) was then computed between the new periodogram realisations, and the CCF was cropped to $\pm$30$\mu$Hz. This value was selected to avoid any contribution from side peaks, the influence of which we would expect to observe at one-half of the large frequency separation, but to comfortably include the expected range of pseudo-mode frequency shifts determined previously ($<$2$\mu$Hz, \citealt{kosak2022multi}). A Lorentzian function was then fitted to the CCF, and the centre of the Lorentzian was recorded as the frequency shift for this iteration. This process was then repeated 100 times. An example of the CCF is shown in Figure \ref{Figure 1}(b), with the fitted Lorentzian shown by the red line. The frequency shift between the two time segments was determined as the mean of the 100 lags, and the shift uncertainty was the standard deviation. This was then repeated for each successive, overlapping time segment over the 22 years of GONG data. 
This results in 70 frequency shifts being calculated between 2001-2023. For later sections (where frequency shifts are compared to GONG magnetogram synoptic maps which start from September 2006 onwards), the magnetic activity proxy is interpolated to 51 frequency shifts between 2006-2023.

\subsection{Latitudinal sensitivity of modes}
    \label{Latitude} 

To identify the temporal pseudo-mode frequencies as a function of latitude, we utilised the ratio described in \citet{simoniello2016new}. For each mode, we calculate $\theta$ as

\begin{equation}\label{eq:theta}
    \theta = \arccos \left( \frac{m}{\sqrt{l(l+1)}} \right) .
\end{equation}

A circle at angle, $\theta$, from the equator passes through \textit{l} number of nodes. Mode sensitivity is greatest at all latitudes below this, and so we use this angle to approximate the upper latitude at which each mode is sensitive.
Modes where $|$\textit{m}$|$=\textit{l}, known as sectoral modes, are most sensitive to regions near the solar equator, and modes where \textit{m}=0, known as zonal modes, are more sensitive to polar regions.

We used this upper latitude to categorise the modes into our defined latitude bands. We then compute a weighted mean to produce the frequency shift and uncertainty for each latitude range.
Pseudo-mode frequency shifts are determined between $0^{\circ} \leq \theta < 75^{\circ}$, as this covers the full latitudinal range over which sunspots drift throughout the solar cycle. From here on, when referring to latitudinal ranges, they should be considered as symmetric about the equator.

\subsection{Magnetic activity proxies}
    \label{Magnetic proxies} 

To compare variation in the temporal pseudo-mode frequency shifts with changes in solar magnetic activity, we employ two proxies.

\subsubsection{$F_{\mathrm{10.7}}$ index}
    \label{F107 index}

The $F_{\mathrm{10.7}}$ index is a global measure of solar magnetic activity \citep{tapping201310}. It quantifies the solar radio flux at a wavelength of 10.7cm, where this emission is integrated over one hour to obtain each measurement (in solar flux units). The $F_{\mathrm{10.7}}$ radio emission originates high in the chromosphere and low in the corona, and has been shown to be well correlated with sunspot number and p-mode frequency shifts \citep{broomhall2015comparison}. The $F_{\mathrm{10.7}}$ index has also been recorded since the mid-20th century, and so data is available for all years that we determined our pseudo-mode frequency shifts measurements for. Throughout this work, the $F_{\mathrm{10.7}}$ index was averaged over 216-day segments (overlapping by 108 days) for ease of comparison to the pseudo-mode frequency shift values determined.

\subsubsection{Magnetic flux density proxy}
    \label{Magnetic flux density proxy} 

To compare variations in the frequencies of the pseudo-modes with the latitudinal progression of the solar cycle, we utilise GONG magnetogram synoptic maps \citep{hill2018global}. Minute by minute observations taken from the six global GONG sites are condensed to create full surface photospheric magnetic flux density maps for each Carrington rotation. Utilising these synoptic maps, the absolute magnetic flux density was determined over selected latitudinal bands.

The categorisation of the modes into each latitude band was determined utilising their upper latitude, which therefore means each mode is sensitive to all latitudes below this (i.e. the mode sensitivity is not confined only within the limits of the latitude band). Therefore, for the magnetic activity proxy, we also consider the full range of values below an upper limit i.e. for frequency shifts calculated for latitude band 30$^\circ$ $\leq$ $\theta$ $<$ 45$^\circ$, we compare against all magnetic flux density data between -45$^\circ$ and 45$^\circ$. 

For each latitudinal band, we calculated the average, absolute magnitude of the magnetic flux density for each synoptic map (one per Carrington rotation). The values are interpolated to the same time frame as the pseudo-mode frequency shifts, and then linearly scaled and shifted for comparison in all figures. The synoptic maps are available from September 2006 onwards.

\begin{figure} 
    \centerline{\includegraphics[width=1\textwidth,clip=]{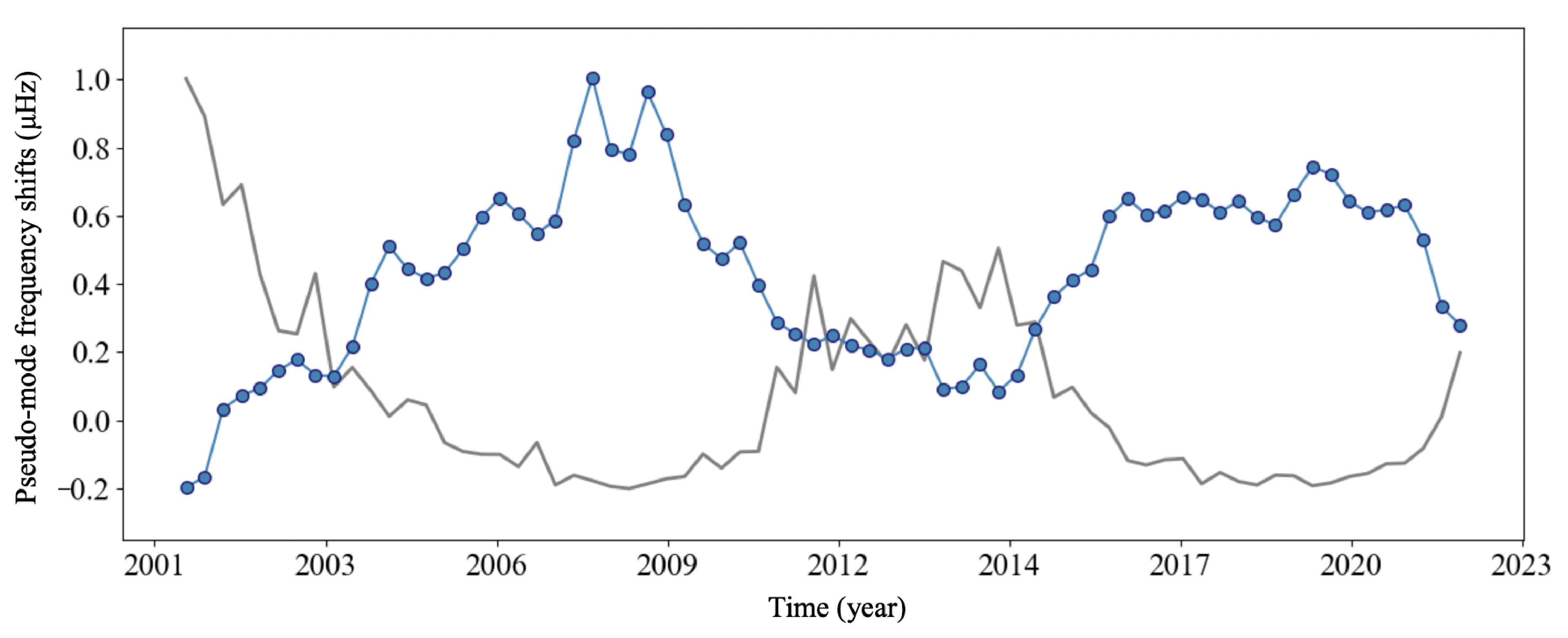}}
    \caption{Solar pseudo-mode frequency shifts (blue data points) averaged over all azimuthal orders, $\textit{m}$, for harmonic degrees 0 $\leq$ $\textit{l}$ $\leq$ 200. Frequency shifts and their uncertainties are generated using the resampled periodogram cross-correlation method with GONG data from 16th June 2001 until 7th June 2023, between time segments of 216 days (overlapping by 108 days). The error bars are too small to be seen, as the shifts are the weighted average of 40\,401 modes. A proxy for magnetic activity, the $F_{\mathrm{10.7}}$ index, is shown by the grey line. The $F_{\mathrm{10.7}}$ index has been linearly re-scaled and shifted for the purpose of this comparison.}
    \label{Figure 2}
\end{figure}

\section{Results}
    \label{Results}

\subsection{Frequency shifts as a function of time}

In our work, we aimed to identify frequency shifts of the solar pseudo-modes as a function of time and latitude across the solar disc using GONG data. To start with, we considered the scenario where a weighted average was taken for all azimuthal orders across all modes (0 $\leq$ $\textit{l}$ $\leq$ 200), and the temporal pseudo-mode frequency shifts over the solar disc as a whole were determined. The results of this are shown by the blue data points in Figure \ref{Figure 2}, where the grey line shows the $F_{\mathrm{10.7}}$ index. The error bars are too small to be seen, as the shifts are the weighted average of 40\,401 modes.

An anti-phase relationship is clearly observed between the pseudo-mode frequency shifts and the $F_{\mathrm{10.7}}$ index throughout the 22 years of GONG data. The Spearman's rank correlation coefficient ($\rho$) identified a very strong negative relationship of -0.93 (p $<$ 10$^{-31}$). 
The frequency shifts were found to vary throughout the timeseries by 1.2$\mu$Hz, an amplitude similar to that observed previously \citep{kosak2022multi}.
In our analysis, we also observe a double peak structure at the frequency shift maxima (corresponding to cycle minima). The peak is most defined at the solar minimum after Cycle 23, but less so for the minimum after Cycle 24 (which was the weakest solar cycle in 100 years). The structure was also observed by \citet{kosak2022multi}, who note the double peak is reminiscent of the double maximum observed in other activity proxies at maxima and has been associated with quasi-biennial oscillations. Overall, our method is able to replicate well the behaviour of the pseudo-mode frequency shifts throughout multiple solar cycles.

We also investigated whether there is any relationship between the GONG duty cycle and the pseudo-mode frequency shifts determined. However, no significant correlation was found ($\rho$=0.03, p=0.8), implying that the magnitude of the pseudo-mode frequency shifts are not impacted by the duty cycle.

\begin{figure} 
    \centerline{\includegraphics[width=0.7\textwidth,clip=]{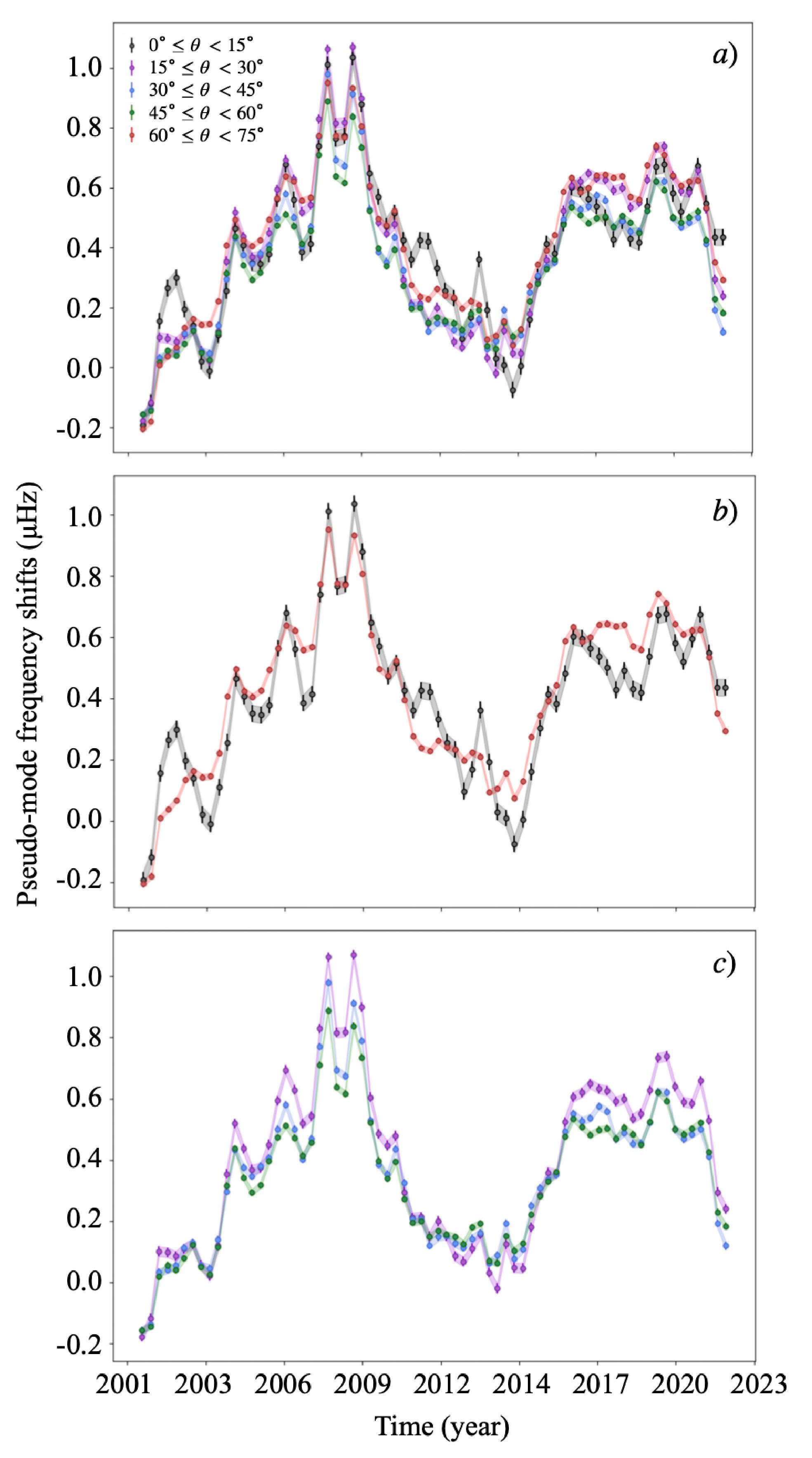}}
    \caption{Frequency shifts of the solar pseudo-modes (5600-6800$\mu$Hz) as a function of time from 16th June 2001 to 7th June 2023. Shifts and their uncertainties are calculated using the resampled periodogram cross-correlation approach. Pseudo-mode frequency shifts across five successive latitude bands over the solar disc are shown, with bands being 15$^\circ$ in size. (a) Displays the frequency shifts for all five latitude bands; (b) shows just the lowest latitudes 0$^\circ$ $\leq$ $\theta$ $<$ 15$^\circ$ (black points) and the highest 60$^\circ$ $\leq$ $\theta$ $<$ 75$^\circ$ (red points); and (c) shows the three mid-latitude bands  15$^\circ$ $\leq$ $\theta$ $<$ 30$^\circ$ (purple points),  30$^\circ$ $\leq$ $\theta$ $<$ 45$^\circ$ (blue points), and  45$^\circ$ $\leq$ $\theta$ $<$ 60$^\circ$ (green points).}
    \label{Figure 3}
\end{figure}

\begin{center}
\begin{table}
    \begin{tabular}{>{\centering\arraybackslash}p{0.35\linewidth}>{\centering\arraybackslash}p{0.35\linewidth}}
        \hline
         Latitude band& Number of modes\\
         \hline
         0$^\circ$ $\leq$ $\theta$ $<$ 15$^\circ$& 1378\\
         15$^\circ$ $\leq$ $\theta$ $<$ 30$^\circ$& 4038\\
         30$^\circ$ $\leq$ $\theta$ $<$ 45$^\circ$& 6416\\
         45$^\circ$ $\leq$ $\theta$ $<$ 60$^\circ$& 8368\\
         60$^\circ$ $\leq$ $\theta$ $<$ 75$^\circ$& 9746\\
       \hline
    \end{tabular}
    \caption{The number of modes used to determine the pseudo-mode frequency shifts for each latitude band.}
    \label{tab:mode_no_15}
\end{table}
\end{center}

\begin{sidewaystable}
\centering
\caption{Pseudo-mode frequency shift variations ($\mu$Hz) across Solar Cycles 23, 24, and the full timeseries. The minimum values of pseudo-mode frequency shifts and their uncertainties for each cycle correspond to the maximum of the solar magnetic cycles. The maximum frequency shift for Solar Cycle 23 corresponds to the solar minimum between Cycle 23 and 24, and the maximum shift for Cycle 24 corresponds to the solar minimum between Cycle 24 and 25. *As Solar Cycle 23 ($\approx$1996-2008) was not fully covered by our analysis, the minimum frequency shift values may be an underestimate.}
\label{Table 1}
\begin{tabular}{lcccccc}
\hline
& \multicolumn{3}{c}{Solar Cycle 23}& \multicolumn{3}{c}{Solar Cycle 24}\\
     Latitude band&   Minimum*& Maximum& Variation& Minimum& Maximum& Variation\\
\hline
0$^\circ$ $\leq$ $\theta$ $<$ 15$^\circ$& -0.19 $\pm$0.02& 1.04 $\pm$0.03& 1.23 $\pm$0.04& -0.08 $\pm$0.03& 0.68 $\pm$0.03& 0.76 $\pm$0.04\\
15$^\circ$ $\leq$ $\theta$ $<$ 30$^\circ$ & -0.18 $\pm$0.01 & 1.07 $\pm$0.03   &1.25 $\pm$0.03 & -0.02 $\pm$0.02&  0.74 $\pm$0.03&  0.76 $\pm$0.04\\
30$^\circ$ $\leq$ $\theta$ $<$ 45$^\circ$ &  -0.16 $\pm$0.01  &0.98 $\pm$0.01  &1.14 $\pm$0.01 & 0.06 $\pm$0.01 & 0.62 $\pm$0.01 &0.56  $\pm$0.01\\
45$^\circ$ $\leq$ $\theta$ $<$ 60$^\circ$& -0.16  $\pm$0.01& 0.89 $\pm$0.01 &1.05 $\pm$0.01& 0.06 $\pm$0.01 & 0.62 $\pm$0.01 &0.56 $\pm$0.01\\
60$^\circ$ $\leq$ $\theta$ $<$ 75$^\circ$& -0.21 $\pm$0.01 &0.95 $\pm$0.01 &1.16  $\pm$0.01& 0.07 $\pm$0.01 &0.74 $\pm$0.01  & 0.67 $\pm$0.01\\
\hline
\end{tabular}
\end{sidewaystable}

\subsection{Frequency shifts as a function of latitude}
    \label{Frequency shifts as a function of latitude}

The variation in the solar pseudo-mode frequency shifts over time as a function of latitude is shown in Figure \ref{Figure 3}, and the number of modes used to calculate the frequency shifts for each latitude band is displayed in Table \ref{tab:mode_no_15}. The results for all five latitude bands (of size 15$^\circ$ covering 0$^\circ$ $\leq$ $\theta$ $<$ 75$^\circ$) are shown in the top panel of Figure \ref{Figure 3}. As expected, for all latitude bands, the pseudo-mode frequency shifts are clearly in anti-phase with the solar magnetic cycle. Frequency shift values are shown to increase whilst magnetic activity declines up until around 2009; fall whilst magnetic activity increases until solar maximum around 2014; and then increase again as solar minimum is approached around 2020.

In all five latitude bands, a double peak structure is observable as frequency shifts reach their maximum values. It is particularly noticeable around the 2009 solar minimum. This double peak structure has previously been observed at solar minimum in pseudo-mode frequency shifts \citep{kosak2022multi}, and at solar maximum in temporal p-mode frequency shifts (\citealt{simoniello2012seismic}; \citealt{simoniello2016new}). The morphology of the shifts also appear to fluctuate regularly throughout the time series with a shorter periodicity than the 11-year cycle. We elaborate further on this in Section \ref{QBO}.

We also observe differences in the morphology of the shifts between the latitude bands. This is best shown in the middle panel of Figure \ref{Figure 3}, where the lowest latitude band 0$^\circ$ $\leq$ $\theta$ $<$ 15$^\circ$ (black points) and the highest 60$^\circ$ $\leq$ $\theta$ $<$ 75$^\circ$ (red points) are shown. From visual inspection, shifts at lower latitudes appear to have greater amplitudes of fluctuation throughout the timeseries, compared to the higher latitudes. 
In addition, the amplitude of the shifts for lower latitudes is greater over the full 22-year timeseries, with lower latitude shifts reaching a larger maximum value and a sharper minimum (1.23~$\pm$0.04~$\mu$Hz for lower, 1.16~$\pm$0.01~$\mu$Hz for higher) (Table~\ref{Table 1}).

The behaviour of the mid-latitude bands is displayed in the last panel of Figure~\ref{Figure 3}, where the pseudo-mode frequency shifts for 15$^\circ$~$\leq$~$\theta$~$<$~30$^\circ$ are shown by the purple data points,  30$^\circ$~$\leq$~$\theta$~$<$~45$^\circ$ by the blue points, and  45$^\circ$~$\leq$~$\theta$~$<$~60$^\circ$ by the green. Overall, the morphology of the shifts is quite similar across these latitudes. The main difference is the amplitude of the frequency shift, where shifts at 15$^\circ$~$\leq$~$\theta$~$<$~30$^\circ$ reach higher values at solar minimum and lower values at solar maximum with an overall amplitude of 1.25~$\pm$0.03~$\mu$Hz (Table~\ref{Table 1}). The total shift variation is smaller at 30$^\circ$~$\leq$~$\theta$~$<$~45$^\circ$ (1.14~$\pm$0.01~$\mu$Hz), and smaller still over 45$^\circ$~$\leq$~$\theta$~$<$~60$^\circ$ (1.05~$\pm$0.01~$\mu$Hz).

\subsection{Comparison to magnetic flux density}
    \label{Comparison to magnetic flux density}

\begin{figure} 
    \centerline{\includegraphics[width=1\textwidth,clip=]{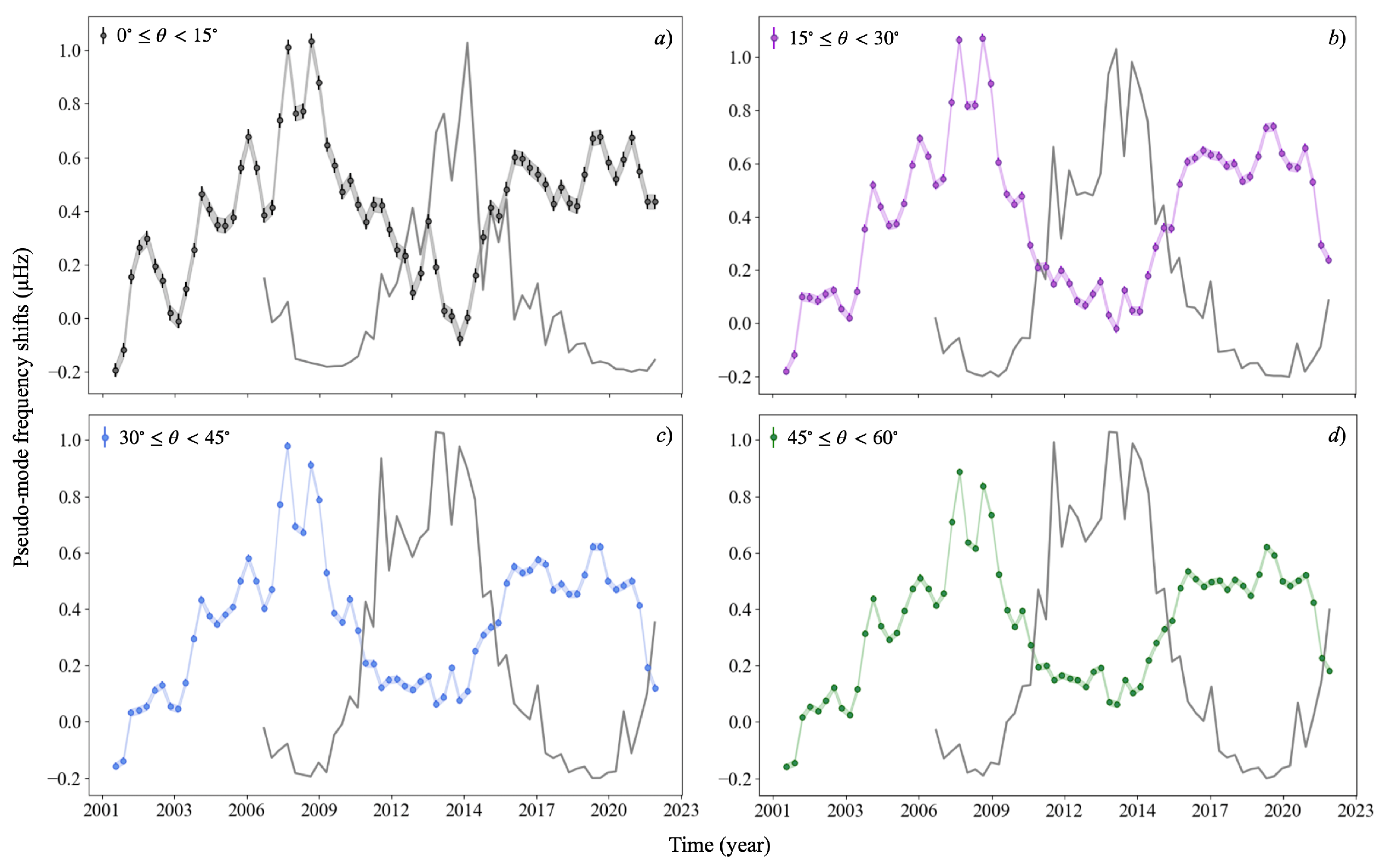}}
    \caption{The solar pseudo-mode frequency shifts as a function of time generated from the resampled periodogram approach. Frequency shifts in (a) for 0$^\circ$ $\leq$ $\theta$ $<$ 15$^\circ$ (black points), in (b) for 15$^\circ$ $\leq$ $\theta$ $<$ 30$^\circ$ (purple points), (c) 30$^\circ$ $\leq$ $\theta$ $<$ 45$^\circ$ (blue points), and (d) 45$^\circ$ $\leq$ $\theta$ $<$ 60$^\circ$ (green points) are shown. The magnetic flux density from GONG synoptic maps (2006 onwards) was interpolated to the same time frame as the frequency shifts, linearly re-scaled, and shifted for comparison, and shown by the grey line on each plot.}
    \label{Figure 4}
\end{figure}

\begin{figure} 
    \centerline{\includegraphics[width=1\textwidth,clip=]{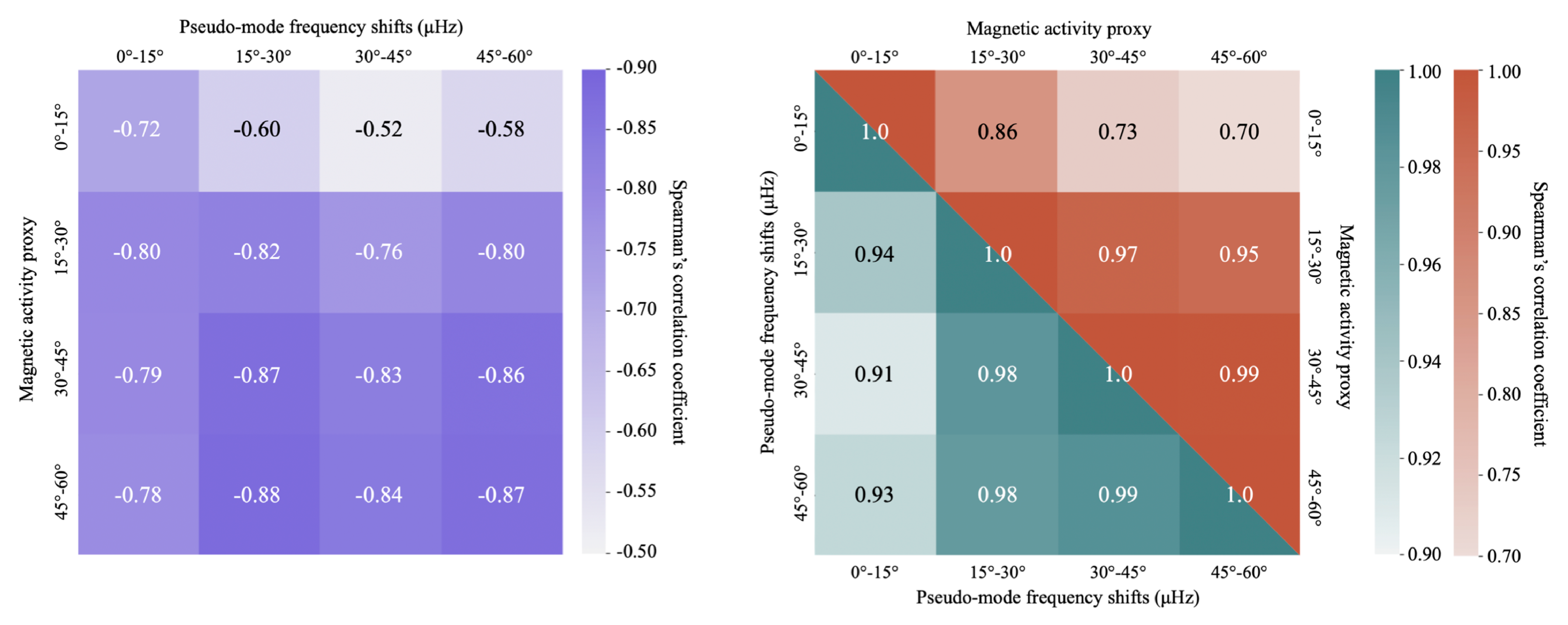}}
    \caption{Left: Spearman's correlation coefficient for the four latitude bands 0$^\circ$ $\leq$ $\theta$ $<$ 15$^\circ$, 15$^\circ$ $\leq$ $\theta$ $<$ 30$^\circ$,  30$^\circ$ $\leq$ $\theta$ $<$ 45$^\circ$, and 45$^\circ$ $\leq$ $\theta$ $<$ 60$^\circ$ comparing the pseudo-mode frequency shifts and the magnetic activity proxy. Right: The correlation between the pseudo-mode frequency shifts against shifts at all latitudes (teal), and the magnetic activity proxy across all latitudes (red).}
    \label{Figure 5}
\end{figure}

In order to compare our latitudinal pseudo-mode frequency shifts with trends in solar magnetic activity over the solar disc, we utilised GONG magnetogram synoptic maps. These maps allowed us to determine absolute, average values for magnetic flux density over various latitude ranges. The results of this comparison are shown in Figure \ref{Figure 4}, where the grey line represents this magnetic activity proxy. Only magnetic flux density data below $\pm$60$^\circ$ is used as data above this latitude is noisier due to line-of-sight projection effects.

The latitudinal drift of sunspots with the progression of the 11-year solar cycle is reflected in the magnetic activity proxy. Initially, sunspots appear at mid-latitudes (around $\pm30^{\circ}$). They then drift closer to the equator, and each cycle concludes with most sunspots around $\pm10^{\circ}$ (which may overlap with the start of the next cycle). This latitudinal drift is clearly visible in Figure \ref{Figure 4} where for latitudes above $\pm15^{\circ}$, the magnetic activity proxy rises sharply in value towards solar maximum. For latitudes below $\pm15^{\circ}$, only a single peak in magnetic activity is observed which appears nearer the end of the solar maximum, reflecting the progression of sunspots to lower latitudes.

Visually, the morphology of the pseudo-mode frequency shifts at all latitudes are clearly in anti-phase with the behaviour of the magnetic activity proxy. The main difference in frequency shift morphology is between latitudes above and below $\pm15^{\circ}$. Below $\pm15^{\circ}$ (Figure \ref{Figure 4}a), the pseudo-mode frequency shifts display a more steady, continuous decline in value to a sharper minimum on the approach to Cycle 24 maximum (from 2009 to 2014). Around 2014, the shift value reflects the single peak behaviour of the magnetic activity proxy, reaching a minimum value at solar maximum.

However, above $\pm15^{\circ}$ (Figure \ref{Figure 4}b, c, and d), pseudo-mode frequency shifts fall in value more suddenly between 2009-2011, and then largely plateau throughout solar maximum until around 2015. This plateau in frequency shift value reflects the continued high values of the magnetic activity proxy between 2011-2015.

The pseudo-mode frequency shift values are also shown to differ between the minimum between Solar Cycles 23 and 24, and the minimum between Cycles 24 and 25, where the value of the frequency shift is greater for the former.
Typically with magnetic activity proxies, the value at cycle minimum is similar between consecutive minima (i.e. for sunspot number, there cannot be less than 0 sunspots). However, as this is not the case with pseudo-mode frequency shifts (which are sensitive to both the strong and weak components of the magnetic field), the difference of the pseudo-mode frequency shifts between solar minima may provide insight into the nature of the weak magnetic field between solar cycles.

The strength of the correlation between the pseudo-mode frequency shifts and the magnetic activity proxy are shown in the correlation matrix in Figure \ref{Figure 5} (see the left hand panel). As expected, at all latitudes, frequency shifts display a significant, strong negative correlation coefficient with magnetic activity proxy. 
To ascertain how frequency shifts for each latitude band progress during the magnetic cycle compared to shifts from other latitudes, we correlate both the pseudo-mode frequency shifts with frequency shifts observed in other latitude bands, and the magnetic activity proxy with the magnetic activity observed at other latitudes (see the right hand panel of Figure \ref{Figure 5}). While we observe changes to frequency shift morphology visually, this correlation better quantifies any agreement between each $15^{\circ}$ band.
When correlating the pseudo-mode frequency shifts with the shifts observed at other latitudes, a lower correlation for frequency shifts in the $0^{\circ} \leq \theta < 15^{\circ}$ band compared to all other latitude bands was found ($\rho$=0.91--0.94). 
In contrast, a higher correlation between the $15^{\circ} \leq \theta < 30^{\circ}$, $30^{\circ} \leq \theta < 45^{\circ}$, and $45^{\circ} \leq \theta < 60^{\circ}$ latitude bands ($\rho$=0.98-0.99) showed that frequency shifts progressed more similarly with time at all latitudes above $\pm15^{\circ}$. 
A similar behaviour was found for magnetic activity proxy, with a weaker correlation below $\pm15^{\circ}$ ($\rho$=0.70-0.86), and a stronger one between all latitude bands above $\pm15^{\circ}$ ($\rho$=0.95-0.99). 
To summarise, both the behaviour of pseudo-mode frequency shifts below $\pm15^{\circ}$, and that of the magnetic activity proxy below $\pm15^{\circ}$, appear to differ compared to all latitude bands above $\pm15^{\circ}$, while above $\pm15^{\circ}$ all latitude bands show a high degree of agreement.

Furthermore, pseudo-mode frequency shifts of higher latitude bands are more strongly correlated with the magnetic activity proxy at the equivalent band (i.e. for $45^{\circ} \leq \theta < 60^{\circ}$, $\rho$=-0.87), compared to lower latitudes (for $0^{\circ} \leq \theta < 15^{\circ}$, $\rho$=-0.72).
The correlation between frequency shifts from the lowest latitude band and the highest latitude magnetic activity proxy ($\rho$=-0.78), is also stronger than the correlation between the highest latitude shifts and lowest latitude magnetic activity proxy ($\rho$=-0.58).
We suggest this is the result of modes still being somewhat sensitive to the magnetic field present at higher latitudes. This occurs because the value of $\theta$ defined by Equation \ref{eq:theta} is not strictly the highest latitude that oscillations are sensitive to. Instead, a mode's sensitively to latitudes greater than $\theta$ is greatly reduced compared to latitudes below $\theta$, but is, importantly, not zero.
In fact, when the magnetic activity proxy is correlated with itself, the correlation is weaker at $0^{\circ} \leq \theta < 15^{\circ}$, reaching $\rho$=0.70. This is to be expected (we know the magnetic activity cycle progresses differently across latitude).
However, when the frequency shifts are correlated with themselves, whilst shifts replicate the behaviour of the magnetic activity proxy in that the correlation is weaker below $\pm15^{\circ}$, the frequency shifts show a higher degree of agreement between each latitude band than the magnetic activity proxy does. Again, this highlights how mode sensitivity is not confined only within the latitude band.

\begin{figure} 
    \centerline{\includegraphics[width=1\textwidth,clip=]{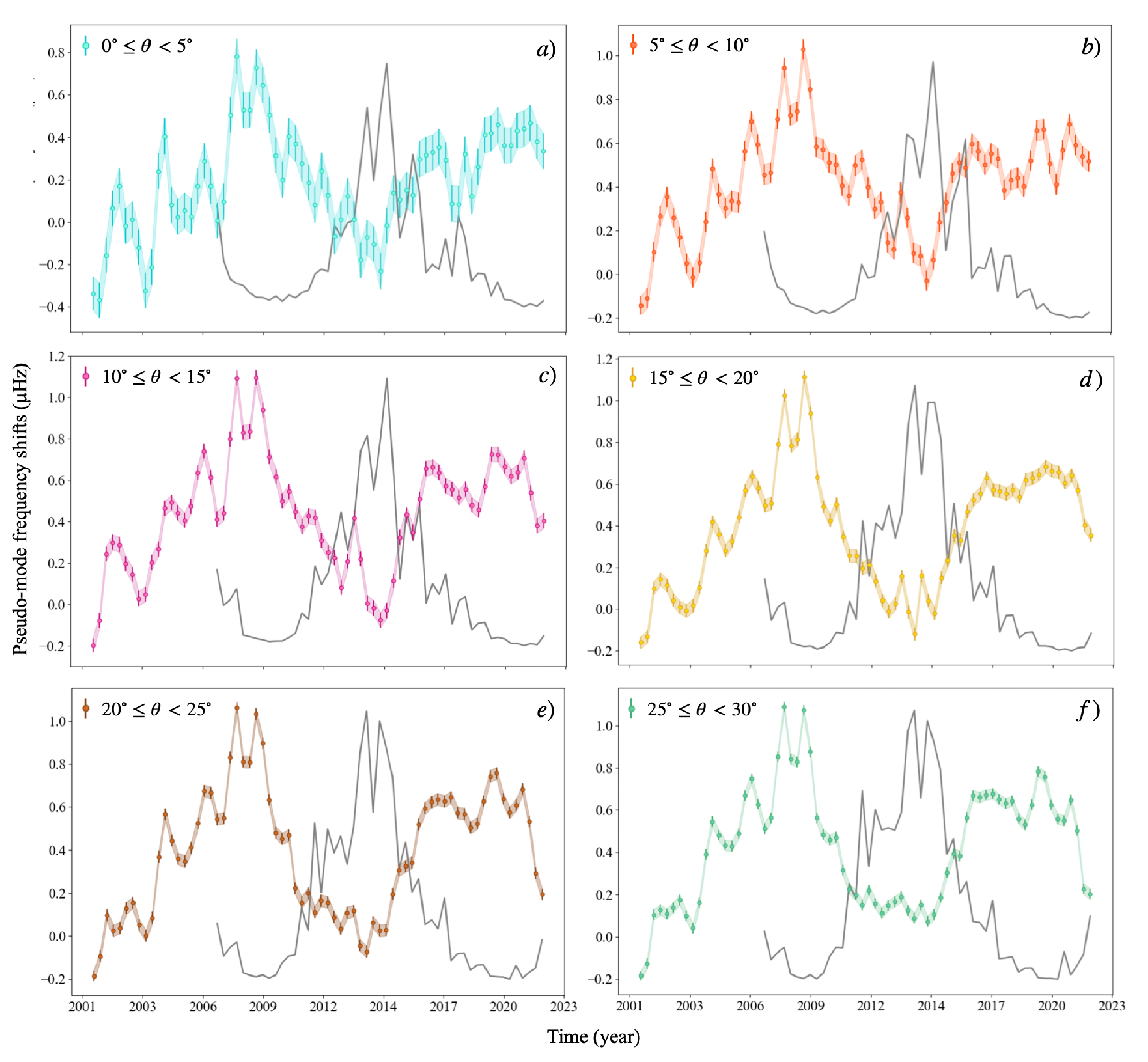}}
    \caption{Pseudo-mode frequency shifts for the Sun using GONG data, calculated by the resampled periodogram method. Shifts and their uncertainties are shown across six latitude bands of size 5$^\circ$ over the solar disc. Frequency shifts in (a) for 0$^\circ$ $\leq$ $\theta$ $<$ 5$^\circ$ (blue points), (b) for 5$^\circ$ $\leq$ $\theta$ $<$ 10$^\circ$ (orange points), (c) 10$^\circ$ $\leq$ $\theta$ $<$ 15$^\circ$ (pink points), (d) 15$^\circ$ $\leq$ $\theta$ $<$ 20$^\circ$ (yellow points), (e) 20$^\circ$ $\leq$ $\theta$ $<$ 25$^\circ$ (brown points),
    and (f) 25$^\circ$ $\leq$ $\theta$ $<$ 30$^\circ$ (green points) are shown. The magnetic flux density from GONG synoptic maps (2006 onwards, grey line) is shown, and has been interpolated to the same time frame as the frequency shifts, linearly re-scaled, and shifted for comparison.}
    \label{Figure 6}
\end{figure}

\begin{figure} 
    \centerline{\includegraphics[width=1\textwidth,clip=]{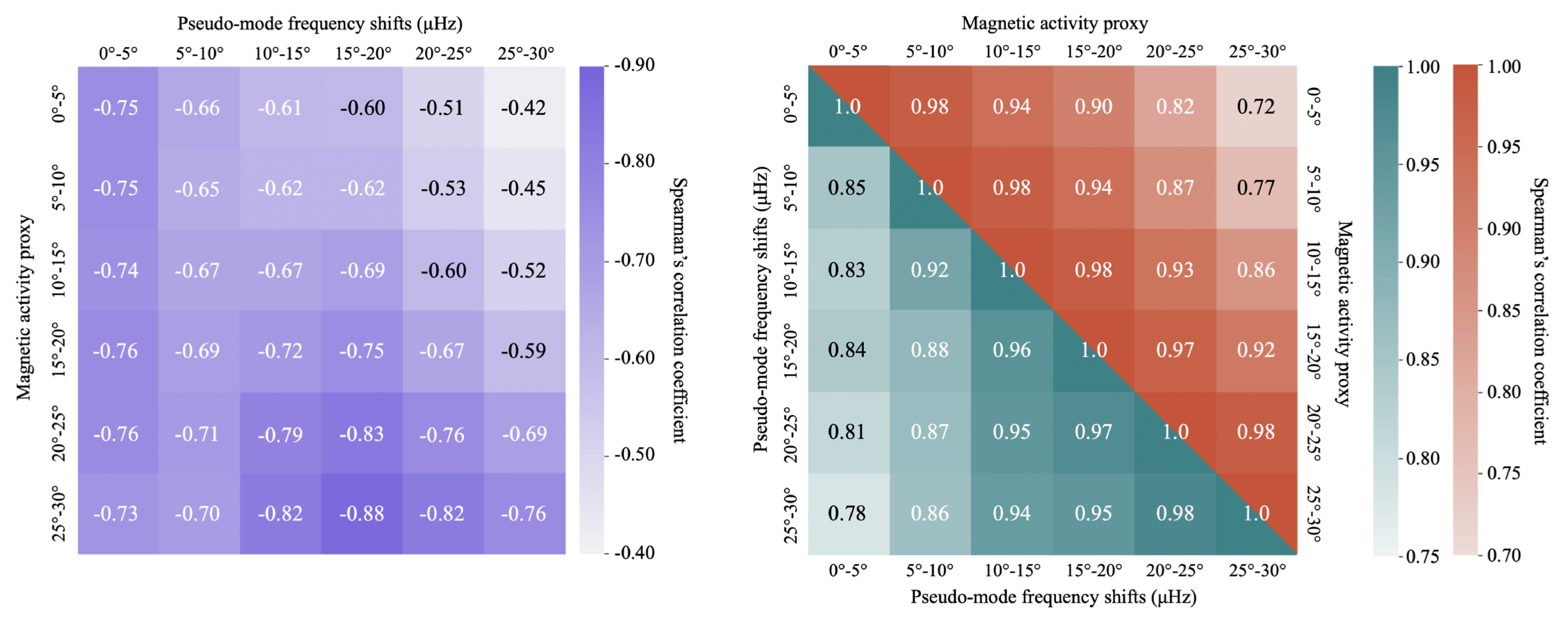}}
    \caption{Left: Spearman's correlation coefficient for the six 5$^\circ$ latitude bands 0$^\circ$ $\leq$ $\theta$ $<$ 5$^\circ$, 5$^\circ$ $\leq$ $\theta$ $<$ 10$^\circ$, 10$^\circ$ $\leq$ $\theta$ $<$ 15$^\circ$,  15$^\circ$ $\leq$ $\theta$ $<$ 20$^\circ$, 20$^\circ$ $\leq$ $\theta$ $<$ 25$^\circ$, and 25$^\circ$ $\leq$ $\theta$ $<$ 30$^\circ$ between the pseudo-mode frequency shifts and the magnetic activity proxy. Right: The correlation between the pseudo-mode frequency shifts against shifts at all latitudes (teal), and the magnetic activity proxy across all latitudes (red).}
    \label{Figure 7}
\end{figure}

\begin{center}
\begin{table}
    \begin{tabular}{>{\centering\arraybackslash}p{0.35\linewidth}>{\centering\arraybackslash}p{0.35\linewidth}}
        \hline
         Latitude band& Number of modes\\
         \hline
         0$^\circ$ $\leq$ $\theta$ $<$ 5$^\circ$& 140\\
         5$^\circ$ $\leq$ $\theta$ $<$ 10$^\circ$& 472\\
         10$^\circ$ $\leq$ $\theta$ $<$ 15$^\circ$& 766\\
         15$^\circ$ $\leq$ $\theta$ $<$ 20$^\circ$& 1058\\
         20$^\circ$ $\leq$ $\theta$ $<$ 25$^\circ$& 1352\\
         25$^\circ$ $\leq$ $\theta$ $<$ 30$^\circ$& 1628\\
       \hline
    \end{tabular}
    \caption{The number of modes used to determine the pseudo-mode frequency shifts for each latitude band.}
    \label{tab:mode_no_5}
\end{table}
\end{center}

\subsection{Frequency shift trends for the lower latitudes} \label{Frequency shift trends for the lower latitudes}
    
As sunspots exist predominantly across low to mid-latitudes on the solar disc, we extended our analysis to focus on changes to pseudo-mode frequency shifts as a function of latitude for smaller latitude bands of $5^{\circ}$ covering 0$^\circ$ $\leq$ $\theta$ $<$ 30$^\circ$. The results of this analysis are shown in Figure \ref{Figure 6}, with the magnetic activity proxy (grey line) for comparison. The error bars are largest for frequency shifts where less modes are included in the weighted average (Table \ref{tab:mode_no_5}).

Again, as expected, all pseudo-mode frequency shifts are anti-correlated with the solar magnetic cycle. As with the results shown in Section \ref{Frequency shifts as a function of latitude}, the morphology of the frequency shifts show a shorter-term (shorter than the 11-year cycle) periodicity. Furthermore, from visual inspection of the frequency shifts, the shape of the temporal variations appears to differ over latitude. A sudden drop in frequency shifts is more apparent (around 2009-2011) in the $20^{\circ} \leq \theta < 25^{\circ}$ and $25^{\circ} \leq \theta < 30^{\circ}$ bands. There is then a plateau in shift values throughout solar maximum. In contrast, the frequency shifts in the lowest latitude bands fall more steadily. Particularly for $0^{\circ} \leq \theta < 5^{\circ}$, the shifts appear to be continually declining at an almost constant rate from 2009 to 2014.

The maximum and minimum value of the frequency shifts also differs. At Solar Cycle 24 maximum, the shifts in $0^{\circ} \leq \theta < 5^{\circ}$ reach a much deeper minimum value of -0.23~$\pm$0.08~$\mu$Hz, where the minimum for $25^{\circ} \leq \theta < 30^{\circ}$ is 0.07~$\pm$0.03~$\mu$Hz. In contrast, at the solar minimum between Cycle 23 and 24, the maximum frequency shift reaches 1.09~$\pm$0.03~$\mu$Hz for the $25^{\circ} \leq \theta < 30^{\circ}$ latitude band, but only 0.78~$\pm$0.08~$\mu$Hz for the $0^{\circ} \leq \theta < 5^{\circ}$ band.

Again, a Spearman's correlation was performed to quantify the strength of the anti-correlation between the pseudo-mode frequency shifts and the magnetic activity proxy for the $5^{\circ}$ latitude bands. The correlation matrix is shown in Figure \ref{Figure 7}. As expected, correlations between pseudo-mode frequency shifts and the magnetic activity proxy are all significantly and strongly negatively correlated.
We also note that the correlation between pseudo-mode frequency shifts at $0^{\circ} \leq \theta < 5^{\circ}$ and the magnetic activity proxy at $25^{\circ} \leq \theta < 30^{\circ}$ are more strongly correlated ($\rho$=-0.73), compared to the correlation between frequency shifts at $25^{\circ} \leq \theta < 30^{\circ}$ and the magnetic activity proxy at $0^{\circ} \leq \theta < 5^{\circ}$ ($\rho$=-0.42). Again, this may be a reflection of the fact that modes categorised into lower latitude bands still have some sensitivity to higher latitudes.
Furthermore, when correlating the pseudo-mode frequency shifts with itself, and the magnetic activity proxy with itself (see the right hand panel of Figure \ref{Figure 7}), we again observe the weakest correlation at lower latitudes for both, highlighting how the temporal progression of the frequency shifts (and magnetic activity) between lower latitudes bands are less similar to those between higher latitude bands. The morphology of pseudo-mode frequency shifts appear to reflect the trends in the magnetic activity proxy, caused by the progression of magnetic activity to lower latitudes.

\subsection{Double peak structure and shorter-term quasi-biennial periodicity}
    \label{QBO}

For all pseudo-mode frequency shifts produced in this work, there appears to be some shorter-term periodicity within the shifts (shorter than the 11-year solar cycle). To further investigate this we utilised wavelet analysis. 

\begin{figure} 
    \centerline{\includegraphics[width=1\textwidth,clip=]{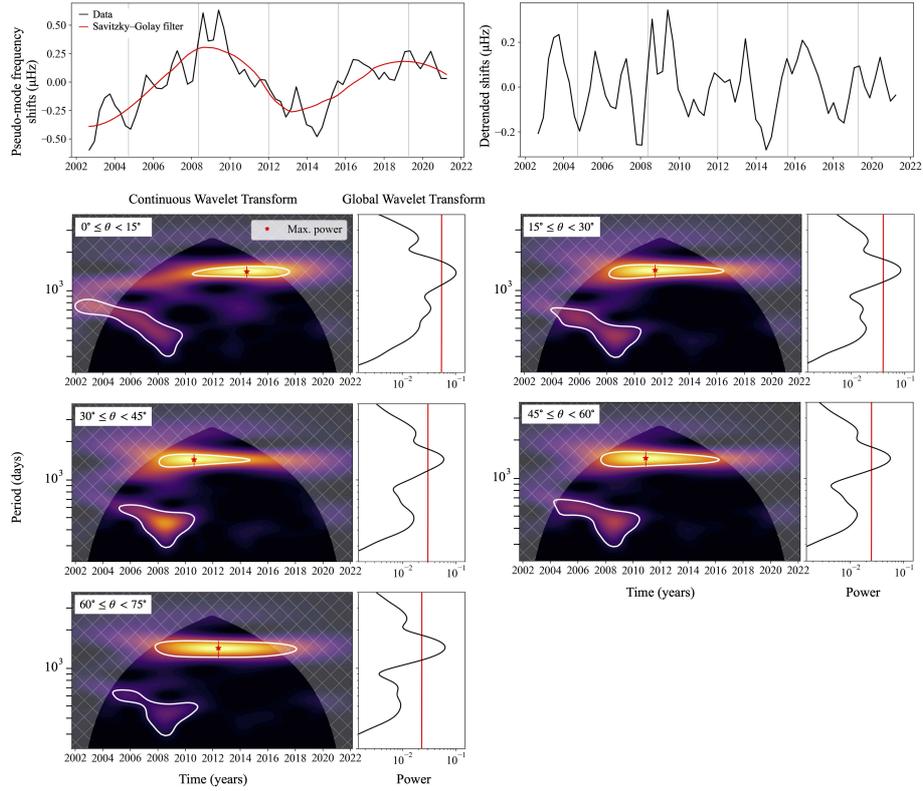}}
    \caption{Top panel (left): Pseudo-mode frequency shifts for the lowest latitude band, 0$^\circ$~$\leq$~$\theta$~$<$~15$^\circ$. The 11-year solar magnetic cycle is clearly visible. A Savitzky–Golay filter (red line) is shown, capturing this 11-year periodicity. Top panel (right): Detrended pseudo-mode frequency shifts with the Savitzky–Golay filter removed, so shorter-term periodicity remains. This is repeated for all five latitude bands analysed by the wavelet analysis, and the continuous wavelet transform (CWT) and global wavelet transform (GWT) are shown below (with latitude bands labelled). Maximum power for each CWT is shown by the red star, and the white contour lines show regions with periodicity at power greater than the 98\% significance level. A 98\% significance level is also shown on the GWT by the red line (which appears continuous, but is lower for shorter periods). All latitude bands have a significant power located between 1417-1450 days (just under four years). The location of maximum power shifts with the solar activity cycle, where at mid-latitudes (30$^\circ$~$\leq$~$\theta$~$<$~45$^\circ$) maximum power exists in August 2010, shifting to lower latitudes (0$^\circ$~$\leq$~$\theta$~$<$~15$^\circ$) by June 2014.}
    \label{Figure 8}
\end{figure}

The analysis is shown in Figure \ref{Figure 8}. The periodicity of the 11-year solar cycle was first removed from the pseudo-mode frequency shifts by smoothing with a Savitzky-Golay filter. A filter with a window length of 21 and a degree 3 polynomial was used as this captured the periodicity of the 11-year solar cycle well, without including any shorter-term periodicity. This 11-year cycle was then removed from the original data. The resultant detrended pseudo-mode frequency shifts are shown in the top panel of Figure \ref{Figure 8}. 
The wavelet analysis was performed using a Morlet wavelet for five latitude bands (of size 15$^\circ$ covering $0^{\circ} \leq \theta < 75^{\circ}$), and the results are shown in the continuous wavelet transform (CWT) heatmap plot for each latitude band. The white contours on the heatmaps are the 98\% confidence contours, and are used to calculate the error on the periodicity at maximum power (by using the minimum and maximum periods within the 98\% confidence contour at the periodicity at maximum power). A global wavelet transform (GWT) is also shown for each latitude band, which displays the sum over time of the power of the periodicities identified in the CWT spectrum. The red line on this plot signifies the 98\% significance level.

For each latitude, we identified a significant periodicity. At the lowest latitude ($0^{\circ} \leq \theta < 15^{\circ}$) the period at maximum power is located at 1417$^{+134}_{-138}$ days. For all other latitudes above $\pm15^{\circ}$, the period at maximum power is the same, at 1450$^{+142}_{-168}$ days for $15^{\circ} \leq \theta < 30^{\circ}$, 1450$^{+139}_{-150}$ days for $30^{\circ} \leq \theta < 45^{\circ}$, 1450$^{+167}_{-196}$ days for $45^{\circ} \leq \theta < 60^{\circ}$, and 1450$^{+183}_{-228}$ days for $60^{\circ} \leq \theta < 75^{\circ}$. This is just under four years for each latitude, and may be a manifestation of the QBO. 
In addition, all heatmaps display a secondary white contoured area between 2002-2010. This region corresponds to a periodicity in the range of 288-850 days (0.8-2.3 years) depending on the latitude band, and, for each band, it is present throughout the decline of Cycle 23 to solar minimum (corresponding to the increase in pseudo-mode frequency shift value). However, the power is not significant above the 98\% significance level on the GWT (shown by the red line). This does not reappear for the decline of Solar Cycle 24 (but may be due to the reduced strength of the latter cycle).

In addition, the location of maximum power of this four-year periodicity varies depending on latitude. At mid-latitudes (30$^\circ$ $\leq$ $\theta$ $<$ 45$^\circ$) maximum power exists during August 2010 (which equates to the rising limb of magnetic activity towards Cycle 24 maximum). However, the power of this four-year periodicity for the lower latitudes (0$^\circ$ $\leq$ $\theta$ $<$ 15$^\circ$) does not reach its maximum until June 2014 (where solar maximum is well established). Like the pseudo-mode frequency shifts at lower latitudes, other solar activity proxies have too been shown to reach maximum QBO power at solar maximum.

\section{Conclusion}
    \label{Conclusion}

In this work, we aimed to identify trends in temporal pseudo-mode frequency shifts as a function of latitude across the solar disc. While previous work on the latitudinal dependence of p-mode frequency shifts has highlighted the potential to better constrain solar dynamo models and magnetic field structures \citep{simoniello2016new}, no analysis, to the best of our knowledge, of the latitudinal behaviour of pseudo-mode frequency shifts currently exists. Here, we utilised GONG data and the resampled periodogram method to calculate temporal pseudo-mode (5600-6800$\mu$Hz) frequency shifts for all azimuthal orders, \textit{m}, of harmonic degrees 0~$\leq$~\textit{l}~$\leq$~200. We then categorised these pseudo-mode frequency shifts into latitude bands, classifying each mode using a ratio to define an upper latitude where mode sensitivity is greatest at latitudes below. Our method was validated by reproducing the strong anti-correlation ($\rho$~=~-0.93, p~$<$~10$^{-31}$) which had previously been found \citep{kosak2022multi} between the weighted average of pseudo-mode frequency shifts (for all \textit{m} over 0 $\leq$ $\textit{l}$ $\leq$ 200) and the $F_{\mathrm{10.7}}$ index as a function of time.

We analysed the morphology of the pseudo-mode frequency shifts for latitude bands of $15^{\circ}$ covering 0$^\circ\leq\theta<75^\circ$. A strong anti-correlation was identified at all latitudes (0$^\circ\leq\theta<60^\circ$) compared to the magnetic flux density determined from GONG synoptic maps. Fluctuations in the shifts were also observed at all latitudes with shorter periodicities than the 11-year magnetic cycle. The morphology of the shifts differed between the lowest and highest latitude bands that was best observed on the rise to Solar Cycle 24 maximum (between 2009-2014). The lowest latitude band (0$^\circ\leq\theta<15^\circ$) had a more gradual, consistent decline in the shift values, whereas the higher latitudes (above $\pm15^\circ$) had a faster drop to minimum, and then a plateau in shift values. This behaviour was reflected in the magnetic activity proxy, with a single peak at solar maximum for latitudes below $\pm15^\circ$ but a double peak for latitudes above $\pm15^\circ$.

Due to the differences between shift morphology above and below $\pm15^\circ$, we focused our analysis on lower latitudes, using latitude bands of $5^{\circ}$ covering 0$^\circ~\leq~\theta~<~30~^\circ$. The pseudo-mode frequency shifts for $0^{\circ} \leq \theta < 5^{\circ}$ reached a sharp minimum value at the maximum of the single peak in the magnetic activity proxy, whereas shifts at $25^{\circ} \leq \theta < 30^{\circ}$ had a faster decline in value and then remained low throughout the double peak in the magnetic activity proxy.

The nature of the temporal pseudo-mode frequency shifts determined here, whereby shifts below $\pm15^\circ$ differ greatly to those above $\pm15^\circ$ (while shifts across each latitude band above $\pm15^\circ$ are in good agreement), is replicated in the magnetic activity proxy. We, therefore, expect the morphological differences in pseudo-mode frequency shifts above and below $\pm15^\circ$ to be a result of the delay in magnetic regions reaching the lower latitudes in each cycle. \citet{vorontsov1998acoustic} modelled the anti-phase variation between pseudo-mode frequencies and solar activity by varying the height of an acoustic potential, which, in turn, impacts the reflection experienced by acoustic oscillations in the solar atmosphere. Assuming such a variation in acoustic potential height is caused by the presence of a magnetic field in the photosphere, it would make sense that the pseudo-mode frequency shifts display the same latitudinal sensitivity as that magnetic activity.

Furthermore, a periodicity shorter than the 11-year solar cycle was identified in the pseudo-mode frequency shifts. To characterise this, we utilised a wavelet analysis for latitude bands of size $15^{\circ}$ covering 0$^\circ$ $\leq$ $\theta$ $<$ 75$^\circ$. For each latitude, a significant (at a 98\% confidence level) periodicity of just under four years was identified. Periodicities shorter than the 11-year solar cycle are well-documented within other solar activity proxies (including the acoustic p modes), and these have been shown to have a range of periods between 0.6-4 years (\citealt{bazilevskaya2014combined}; \citealt{mehta2022cycle}; \citealt{jain2023helioseismic}).

The behaviour of pseudo-mode frequency shifts has previously been shown to reflect trends (by moving in anti-phase) in solar magnetic activity. We aimed to extend this to search for a latitudinal dependence to pseudo-mode frequency shifts, to ascertain whether there is any difference in pseudo-mode shifts and amplitudes across the solar disc through multiple magnetic activity cycles. In doing so, we also identified shorter four-year long periodicities. Our work emphasises the use and contribution of high frequencies beyond the acoustic cut-off in our understanding of the latitudinal progression of the 11-year solar cycle, and the variable nature of shorter activity cycles.

%%%%%%%%%%%%%%%%%%%%%%%%%%%%%%%%%%%%%%%%%%%%%%%%%%%%%%%%%%%%%%%%%%%%%%%%%%%
%% Appendix
%
% \appendix   

%%%%%%%%%%%%%%%%%%%%%%%%%%%%%%%%%%%%%%%%%%%%%%%%%%%%%%%%%%%%%%%%%%%%%%%%%%%
%% Acknowledgements

\begin{acks}
We gratefully acknowledge support from the UK Science and Technology Facilities Council (STFC) grant ST/W507908/1. A-M. Broomhall acknowledges support from STFC grant ST/X000915/1, and A-M. Broomhall and T. Mehta acknowledge support from STFC grant ST/T000252/1.

This study also acknowledges use of Python packages \texttt{NumPy} \citep{harris2020array}, \texttt{Matplotlib} \citep{Hunter2007}, \texttt{SciPy} \citep{2020SciPy-NMeth}, and the Python \texttt{multiprocessing} library. This work also made use of \texttt{AstroPy}, a community-developed core Python package and an ecosystem of tools and resources for astronomy \citep{astropy:2013, astropy:2018, astropy:2022}. Wavelet analysis software was provided by T. Mehta is available at \url{https://github.com/TishtryaMehta/QBO_evolution}. This work utilises data from the National Solar Observatory Integrated Synoptic Program, which is operated by the Association of Universities for Research in Astronomy, under a cooperative agreement with the National Science Foundation and with additional financial support from the National Oceanic and Atmospheric Administration, the National Aeronautics and Space Administration, and the United States Air Force. The GONG network of instruments is hosted by the Big Bear Solar Observatory, High Altitude Observatory, Learmonth Solar Observatory, Udaipur Solar Observatory, Instituto de Astrofísica de Canarias, and Cerro Tololo Interamerican Observatory. 
\end{acks}

%% Available additional data environments:
%% required: authorcontribution, fundinginformation, dataavailability

\begin{authorcontribution}
L. J. Millson conducted the main data analysis and wrote the manuscript and code used. A-M. Broomhall conceptualised the research aims and reviewed the manuscript. T. Mehta provided the wavelet analysis code used in Section \ref{QBO}.
\end{authorcontribution}

\begin{fundinginformation}
This research was supported by the UK Science and Technology Facilities Council (STFC).
\end{fundinginformation}

\begin{dataavailability}
All data used in this study is publicly available. $F_{\mathrm{10.7}}$ index data is available online at \url{ftp://ftp.seismo.nrcan.gc.ca/spaceweather/solar flux/}). Network merged timeseries (mrvmt) and synoptic maps from the Global Oscillations Network Group (GONG) are available at \url{https://gong.nso.edu}).
\end{dataavailability}

% \begin{ethics}
% \begin{conflict}

% \end{conflict}
% \end{ethics}

%%% %%%%%%%%%%%%%%%%%%%%%%%%%%%%%%%%%%%%%%%%%%%%%%%%%%%%%%%%%%%
%% Bibliography

% Using BibTeX

\bibliographystyle{spr-mp-sola}
\bibliography{main_bibliography.bib}

\begin{thebibliography}{50}
% BibTex style file: spr-mp-sola.bst, v2.06, 2023-07-27
\ifx\bisbn     \undefined \def\bisbn  #1{ISBN #1}\fi
\ifx\binits    \undefined \def\binits#1{#1}\fi
\ifx\bauthor   \undefined \def\bauthor#1{#1}\fi
\ifx\batitle   \undefined \def\batitle#1{#1}\fi
\ifx\bjtitle   \undefined \def\bjtitle#1{\textit{#1}}\fi
\ifx\bvolume   \undefined \def\bvolume#1{\textbf{#1}}\fi
\ifx\byear     \undefined \def\byear#1{#1}\fi
\ifx\bissue    \undefined \def\bissue#1{#1}\fi
\ifx\bfpage    \undefined \def\bfpage#1{#1}\fi
\ifx\blpage    \undefined \def\blpage #1{#1}\fi
\ifx\burl      \undefined \def\burl#1{#1}\fi
\ifx\href      \undefined \def\href#1#2{#2}\fi
\ifx\betal     \undefined \def\betal{et al.}\fi
\ifx\bctitle   \undefined \def\bctitle#1{#1}\fi
\ifx\beditor   \undefined \def\beditor#1{#1}\fi
\ifx\bbtitle   \undefined \def\bbtitle#1{\textit{#1}}\fi
\ifx\bedition  \undefined \def\bedition#1{#1}\fi
\ifx\bseriesno \undefined \def\bseriesno#1{\textbf{#1}}\fi
\ifx\blocation \undefined \def\blocation#1{#1}\fi
\ifx\bsertitle \undefined \def\bsertitle#1{\textit{#1}}\fi
\ifx\bsnm      \undefined \def\bsnm#1{#1}\fi
\ifx\bsuffix   \undefined \def\bsuffix#1{#1}\fi
\ifx\bparticle \undefined \def\bparticle#1{#1}\fi
\ifx\barticle  \undefined \def\barticle#1{}\fi
\ifx\binstitute  \undefined \def\binstitute#1{#1}\fi
\ifx\bpublisher  \undefined \def\bpublisher#1{#1}\fi
\ifx\doiurl    \undefined \def\doiurl#1{\href{#1}{DOI}}\fi
\makeatletter
\def\safeHref#1#2#3{\in@{http}{#2}\ifin@\href{#2}{#3}\else\href{#1#2}{#3}\fi}
\makeatother
\ifx\adsurl    \undefined \def\adsurl#1{\safeHref{https://ui.adsabs.harvard.edu/abs/}{#1}{ADS}}\fi
\ifx\arxivurl  \undefined \def\arxivurl#1{\safeHref{http://arxiv.org/abs/}{#1}{arXiv}}\fi
\ifx\botherref \undefined \def\botherref#1{}\fi
\ifx\url       \undefined \def\url#1{#1}\fi
\ifx\bchapter  \undefined \def\bchapter#1{}\fi
\ifx\bbook     \undefined \def\bbook#1{}\fi
\ifx\bcomment  \undefined \def\bcomment#1{#1}\fi
\ifx\oauthor   \undefined \def\oauthor#1{#1}\fi
\ifx\citeauthoryear \undefined\def \citeauthoryear#1{#1}\fi
\def\endbibitem {}
\ifx\bconflocation  \undefined \def\bconflocation#1{#1} \fi

\bibitem[\protect\citeauthoryear{Anguera~Gubau et~al.}{1992}]{anguera1992low}
\begin{barticle}
\bauthor{\bsnm{Anguera~Gubau}, \binits{M.}},
\bauthor{\bsnm{Pall{\'e}}, \binits{P.}},
\bauthor{\bsnm{Perez~Hernandez}, \binits{F.}},
\bauthor{\bsnm{R{\'e}gulo}, \binits{C.}},
\bauthor{\bsnm{Roca~Cort{\'e}s}, \binits{T.}}:
\byear{1992},
\batitle{The low l solar p-mode spectrum at maximum and minimum solar activity}.
\bjtitle{Astronomy and Astrophysics}
\bvolume{255},
\bfpage{363}.
\end{barticle}
\endbibitem

\bibitem[\protect\citeauthoryear{{Astropy Collaboration}}{2013}]{astropy:2013}
\begin{barticle}
\bauthor{\bsnm{{Astropy Collaboration}}}:
\byear{2013},
\batitle{{Astropy: A community Python package for astronomy}}.
\bjtitle{Astronomy and Astrophysics}
\bvolume{558},
\bfpage{A33}.
\end{barticle}
\endbibitem

\bibitem[\protect\citeauthoryear{{Astropy Collaboration}}{2018}]{astropy:2018}
\begin{barticle}
\bauthor{\bsnm{{Astropy Collaboration}}}:
\byear{2018},
\batitle{{The Astropy Project: Building an Open-science Project and Status of the v2.0 Core Package}}.
\bjtitle{The Astronomical Journal}
\bvolume{156},
\bfpage{123}.
\end{barticle}
\endbibitem

\bibitem[\protect\citeauthoryear{{Astropy Collaboration}}{2022}]{astropy:2022}
\begin{barticle}
\bauthor{\bsnm{{Astropy Collaboration}}}:
\byear{2022},
\batitle{{The Astropy Project: Sustaining and Growing a Community-oriented Open-source Project and the Latest Major Release (v5.0) of the Core Package}}.
\bjtitle{Astrophysical Journal}
\bvolume{935},
\bfpage{167}.
\end{barticle}
\endbibitem

\bibitem[\protect\citeauthoryear{Basu}{2016}]{basu2016global}
\begin{barticle}
\bauthor{\bsnm{Basu}, \binits{S.}}:
\byear{2016},
\batitle{Global seismology of the Sun}.
\bjtitle{Living Reviews in Solar Physics}
\bvolume{13},
\bfpage{2}.
\end{barticle}
\endbibitem

\bibitem[\protect\citeauthoryear{{Basu} et~al.}{2012}]{2012ApJ...758...43B}
\begin{barticle}
\bauthor{\bsnm{{Basu}}, \binits{S.}},
\bauthor{\bsnm{{Broomhall}}, \binits{A.-M.}},
\bauthor{\bsnm{{Chaplin}}, \binits{W.J.}},
\bauthor{\bsnm{{Elsworth}}, \binits{Y.}}:
\byear{2012},
\batitle{{Thinning of the Sun's Magnetic Layer: The Peculiar Solar Minimum Could Have Been Predicted}}.
\bjtitle{The Astrophysical Journal}
\bvolume{758},
\bfpage{43}.
\end{barticle}
\endbibitem

\bibitem[\protect\citeauthoryear{Bazilevskaya et~al.}{2014}]{bazilevskaya2014combined}
\begin{barticle}
\bauthor{\bsnm{Bazilevskaya}, \binits{G.}},
\bauthor{\bsnm{Broomhall}, \binits{A.-M.}},
\bauthor{\bsnm{Elsworth}, \binits{Y.}},
\bauthor{\bsnm{Nakariakov}, \binits{V.}}:
\byear{2014},
\batitle{A combined analysis of the observational aspects of the quasi-biennial oscillation in solar magnetic activity}.
\bjtitle{Space Science Reviews}
\bvolume{186},
\bfpage{359}.
\end{barticle}
\endbibitem

\bibitem[\protect\citeauthoryear{{Broomhall}}{2017}]{2017SoPh..292...67B}
\begin{barticle}
\bauthor{\bsnm{{Broomhall}}, \binits{A.-M.}}:
\byear{2017},
\batitle{{A Helioseismic Perspective on the Depth of the Minimum Between Solar Cycles 23 and 24}}.
\bjtitle{Solar Physics}
\bvolume{292},
\bfpage{67}.
\end{barticle}
\endbibitem

\bibitem[\protect\citeauthoryear{Broomhall and Nakariakov}{2015}]{broomhall2015comparison}
\begin{barticle}
\bauthor{\bsnm{Broomhall}, \binits{A.-M.}},
\bauthor{\bsnm{Nakariakov}, \binits{V.M.}}:
\byear{2015},
\batitle{A comparison between global proxies of the Sun’s magnetic activity cycle: Inferences from helioseismology}.
\bjtitle{Solar Physics}
\bvolume{290},
\bfpage{3095}.
\end{barticle}
\endbibitem

\bibitem[\protect\citeauthoryear{Broomhall et~al.}{2009}]{broomhall2009current}
\begin{barticle}
\bauthor{\bsnm{Broomhall}, \binits{A.-M.}},
\bauthor{\bsnm{Chaplin}, \binits{W.}},
\bauthor{\bsnm{Elsworth}, \binits{Y.}},
\bauthor{\bsnm{Fletcher}, \binits{S.}},
\bauthor{\bsnm{New}, \binits{R.}}:
\byear{2009},
\batitle{Is the current lack of solar activity only skin deep?}
\bjtitle{The Astrophysical Journal}
\bvolume{700},
\bfpage{L162}.
\end{barticle}
\endbibitem

\bibitem[\protect\citeauthoryear{Broomhall et~al.}{2011}]{broomhall2011short}
\begin{bchapter}
\bauthor{\bsnm{Broomhall}, \binits{A.-M.}},
\bauthor{\bsnm{Fletcher}, \binits{S.T.}},
\bauthor{\bsnm{Salabert}, \binits{D.}},
\bauthor{\bsnm{Basu}, \binits{S.}},
\bauthor{\bsnm{Chaplin}, \binits{W.J.}},
\bauthor{\bsnm{Elsworth}, \binits{Y.}},
\bauthor{\bsnm{Garc{\'\i}a}, \binits{R.A.}},
\bauthor{\bsnm{Jim{\'e}nez}, \binits{A.}},
\bauthor{\bsnm{New}, \binits{R.}}:
\byear{2011},
\bctitle{Are short-term variations in solar oscillation frequencies the signature of a second solar dynamo?}
In: \bbtitle{Journal of Physics: Conference Series}
\bseriesno{271}.
\end{bchapter}
\endbibitem

\bibitem[\protect\citeauthoryear{Broomhall et~al.}{2012}]{broomhall2012quasi}
\begin{barticle}
\bauthor{\bsnm{Broomhall}, \binits{A.-M.}},
\bauthor{\bsnm{Chaplin}, \binits{W.}},
\bauthor{\bsnm{Elsworth}, \binits{Y.}},
\bauthor{\bsnm{Simoniello}, \binits{R.}}:
\byear{2012},
\batitle{Quasi-biennial variations in helioseismic frequencies: can the source of the variation be localized?}
\bjtitle{Monthly Notices of the Royal Astronomical Society}
\bvolume{420},
\bfpage{1405}.
\end{barticle}
\endbibitem

\bibitem[\protect\citeauthoryear{Broomhall et~al.}{2014}]{broomhall2014sun}
\begin{barticle}
\bauthor{\bsnm{Broomhall}, \binits{A.-M.}},
\bauthor{\bsnm{Chatterjee}, \binits{P.}},
\bauthor{\bsnm{Howe}, \binits{R.}},
\bauthor{\bsnm{Norton}, \binits{A.}},
\bauthor{\bsnm{Thompson}, \binits{M.}}:
\byear{2014},
\batitle{The Sun’s interior structure and dynamics, and the solar cycle}.
\bjtitle{Space Science Reviews}
\bvolume{186},
\bfpage{191}.
\end{barticle}
\endbibitem

\bibitem[\protect\citeauthoryear{Chaplin et~al.}{2000}]{chaplin2000variations}
\begin{barticle}
\bauthor{\bsnm{Chaplin}, \binits{W.}},
\bauthor{\bsnm{Elsworth}, \binits{Y.}},
\bauthor{\bsnm{Isaak}, \binits{G.}},
\bauthor{\bsnm{Miller}, \binits{B.}},
\bauthor{\bsnm{New}, \binits{R.}}:
\byear{2000},
\batitle{Variations in the excitation and damping of low-$\ell$ solar p modes over the solar activity cycle}.
\bjtitle{Monthly Notices of the Royal Astronomical Society}
\bvolume{313},
\bfpage{32}.
\end{barticle}
\endbibitem

\bibitem[\protect\citeauthoryear{Duvall~Jr et~al.}{1991}]{duvall1991measurements}
\begin{barticle}
\bauthor{\bsnm{Duvall~Jr}, \binits{T.}},
\bauthor{\bsnm{Harvey}, \binits{J.}},
\bauthor{\bsnm{Jefferies}, \binits{S.}},
\bauthor{\bsnm{Pomerantz}, \binits{M.}}:
\byear{1991},
\batitle{Measurements of high-frequency solar oscillation modes}.
\bjtitle{Astrophysical Journal}
\bvolume{373},
\bfpage{308}.
\end{barticle}
\endbibitem

\bibitem[\protect\citeauthoryear{Elsworth et~al.}{1990}]{elsworth1990variation}
\begin{barticle}
\bauthor{\bsnm{Elsworth}, \binits{Y.}},
\bauthor{\bsnm{Howe}, \binits{R.}},
\bauthor{\bsnm{Isaak}, \binits{G.}},
\bauthor{\bsnm{McLeod}, \binits{C.}},
\bauthor{\bsnm{New}, \binits{R.}}:
\byear{1990},
\batitle{Variation of low-order acoustic solar oscillations over the solar cycle}.
\bjtitle{Nature}
\bvolume{345},
\bfpage{322}.
\end{barticle}
\endbibitem

\bibitem[\protect\citeauthoryear{Fletcher et~al.}{2010}]{fletcher2010seismic}
\begin{barticle}
\bauthor{\bsnm{Fletcher}, \binits{S.T.}},
\bauthor{\bsnm{Broomhall}, \binits{A.-M.}},
\bauthor{\bsnm{Salabert}, \binits{D.}},
\bauthor{\bsnm{Basu}, \binits{S.}},
\bauthor{\bsnm{Chaplin}, \binits{W.J.}},
\bauthor{\bsnm{Elsworth}, \binits{Y.}},
\bauthor{\bsnm{Garcia}, \binits{R.A.}},
\bauthor{\bsnm{New}, \binits{R.}}:
\byear{2010},
\batitle{A seismic signature of a second dynamo?}
\bjtitle{The Astrophysical Journal Letters}
\bvolume{718},
\bfpage{L19}.
\end{barticle}
\endbibitem

\bibitem[\protect\citeauthoryear{Garc{\'\i}a et~al.}{1998}]{garcia1998high}
\begin{barticle}
\bauthor{\bsnm{Garc{\'\i}a}, \binits{R.}},
\bauthor{\bsnm{Pall{\'e}}, \binits{P.}},
\bauthor{\bsnm{Turck-Chieze}, \binits{S.}},
\bauthor{\bsnm{Osaki}, \binits{Y.}},
\bauthor{\bsnm{Shibahashi}, \binits{H.}},
\bauthor{\bsnm{Jefferies}, \binits{S.}},
\bauthor{\bsnm{Boumier}, \binits{P.}},
\bauthor{\bsnm{Gabriel}, \binits{A.}},
\bauthor{\bsnm{Grec}, \binits{G.}},
\bauthor{\bsnm{Robillot}, \binits{J.}}, \betal:
\byear{1998},
\batitle{High-frequency peaks in the power spectrum of solar velocity observations from the GOLF experiment}.
\bjtitle{The Astrophysical Journal}
\bvolume{504},
\bfpage{L51}.
\end{barticle}
\endbibitem

\bibitem[\protect\citeauthoryear{Goldreich et~al.}{1991}]{goldreich1991implications}
\begin{barticle}
\bauthor{\bsnm{Goldreich}, \binits{P.}},
\bauthor{\bsnm{Murray}, \binits{N.}},
\bauthor{\bsnm{Willette}, \binits{G.}},
\bauthor{\bsnm{Kumar}, \binits{P.}}:
\byear{1991},
\batitle{Implications of solar p-mode frequency shifts}.
\bjtitle{Astrophysical Journal}
\bvolume{370},
\bfpage{752}.
\end{barticle}
\endbibitem

\bibitem[\protect\citeauthoryear{Harris et~al.}{2020}]{harris2020array}
\begin{barticle}
\bauthor{\bsnm{Harris}, \binits{C.R.}},
\bauthor{\bsnm{Millman}, \binits{K.J.}},
\bauthor{\bparticle{van~der} \bsnm{Walt}, \binits{S.J.}},
\bauthor{\bsnm{Gommers}, \binits{R.}},
\bauthor{\bsnm{Virtanen}, \binits{P.}},
\bauthor{\bsnm{Cournapeau}, \binits{D.}},
\bauthor{\bsnm{Wieser}, \binits{E.}},
\bauthor{\bsnm{Taylor}, \binits{J.}},
\bauthor{\bsnm{Berg}, \binits{S.}},
\bauthor{\bsnm{Smith}, \binits{N.J.}},
\bauthor{\bsnm{Kern}, \binits{R.}},
\bauthor{\bsnm{Picus}, \binits{M.}},
\bauthor{\bsnm{Hoyer}, \binits{S.}},
\bauthor{\bparticle{van} \bsnm{Kerkwijk}, \binits{M.H.}},
\bauthor{\bsnm{Brett}, \binits{M.}},
\bauthor{\bsnm{Haldane}, \binits{A.}},
\bauthor{\bparticle{del} \bsnm{R{\'{i}}o}, \binits{J.F.}},
\bauthor{\bsnm{Wiebe}, \binits{M.}},
\bauthor{\bsnm{Peterson}, \binits{P.}},
\bauthor{\bsnm{G{\'{e}}rard-Marchant}, \binits{P.}},
\bauthor{\bsnm{Sheppard}, \binits{K.}},
\bauthor{\bsnm{Reddy}, \binits{T.}},
\bauthor{\bsnm{Weckesser}, \binits{W.}},
\bauthor{\bsnm{Abbasi}, \binits{H.}},
\bauthor{\bsnm{Gohlke}, \binits{C.}},
\bauthor{\bsnm{Oliphant}, \binits{T.E.}}:
\byear{2020},
\batitle{Array programming with {NumPy}}.
\bjtitle{Nature}
\bvolume{585},
\bfpage{357}.
\end{barticle}
\endbibitem

\bibitem[\protect\citeauthoryear{Harvey et~al.}{1996}]{harvey1996global}
\begin{barticle}
\bauthor{\bsnm{Harvey}, \binits{J.}},
\bauthor{\bsnm{Hill}, \binits{F.}},
\bauthor{\bsnm{Hubbard}, \binits{R.}},
\bauthor{\bsnm{Kennedy}, \binits{J.}},
\bauthor{\bsnm{Leibacher}, \binits{J.}},
\bauthor{\bsnm{Pintar}, \binits{J.}},
\bauthor{\bsnm{Gilman}, \binits{P.}},
\bauthor{\bsnm{Noyes}, \binits{R.}},
\bauthor{\bsnm{Title}, \binits{A.}},
\bauthor{\bsnm{Toomre}, \binits{J.}}, \betal:
\byear{1996},
\batitle{The global oscillation network group (GONG) project}.
\bjtitle{Science}
\bvolume{272},
\bfpage{1284}.
\end{barticle}
\endbibitem

\bibitem[\protect\citeauthoryear{Hill}{2018}]{hill2018global}
\begin{barticle}
\bauthor{\bsnm{Hill}, \binits{F.}}:
\byear{2018},
\batitle{The global oscillation network group facility - an example of research to operations in space weather}.
\bjtitle{Space Weather}
\bvolume{16},
\bfpage{1488}.
\end{barticle}
\endbibitem

\bibitem[\protect\citeauthoryear{Hill et~al.}{1996}]{hill1996solar}
\begin{barticle}
\bauthor{\bsnm{Hill}, \binits{F.}},
\bauthor{\bsnm{Stark}, \binits{P.}},
\bauthor{\bsnm{Stebbins}, \binits{R.}},
\bauthor{\bsnm{Anderson}, \binits{E.}},
\bauthor{\bsnm{Antia}, \binits{H.}},
\bauthor{\bsnm{Brown}, \binits{T.}},
\bauthor{\bsnm{Duvall~Jr}, \binits{T.}},
\bauthor{\bsnm{Haber}, \binits{D.}},
\bauthor{\bsnm{Harvey}, \binits{J.}},
\bauthor{\bsnm{Hathaway}, \binits{D.}}, \betal:
\byear{1996},
\batitle{The solar acoustic spectrum and eigenmode parameters}.
\bjtitle{Science}
\bvolume{272},
\bfpage{1292}.
\end{barticle}
\endbibitem

\bibitem[\protect\citeauthoryear{Howe, Komm, and Hill}{2002}]{howe2002localizing}
\begin{barticle}
\bauthor{\bsnm{Howe}, \binits{R.}},
\bauthor{\bsnm{Komm}, \binits{R.}},
\bauthor{\bsnm{Hill}, \binits{F.}}:
\byear{2002},
\batitle{Localizing the solar cycle frequency shifts in global p-modes}.
\bjtitle{The Astrophysical Journal}
\bvolume{580},
\bfpage{1172}.
\end{barticle}
\endbibitem

\bibitem[\protect\citeauthoryear{{Howe} et~al.}{2018}]{2018MNRAS.480L..79H}
\begin{barticle}
\bauthor{\bsnm{{Howe}}, \binits{R.}},
\bauthor{\bsnm{{Chaplin}}, \binits{W.J.}},
\bauthor{\bsnm{{Davies}}, \binits{G.R.}},
\bauthor{\bsnm{{Elsworth}}, \binits{Y.}},
\bauthor{\bsnm{{Basu}}, \binits{S.}},
\bauthor{\bsnm{{Broomhall}}, \binits{A.-M.}}:
\byear{2018},
\batitle{{Changes in the sensitivity of solar p-mode frequency shifts to activity over three solar cycles}}.
\bjtitle{Monthly Notices of the Royal Astronomical Society}
\bvolume{480},
\bfpage{L79}.
\end{barticle}
\endbibitem

\bibitem[\protect\citeauthoryear{Hunter}{2007}]{Hunter2007}
\begin{barticle}
\bauthor{\bsnm{Hunter}, \binits{J.D.}}:
\byear{2007},
\batitle{Matplotlib: A 2D graphics environment}.
\bjtitle{Computing in Science \& Engineering}
\bvolume{9},
\bfpage{90}.
\end{barticle}
\endbibitem

\bibitem[\protect\citeauthoryear{Jain, Chowdhury, and Tripathy}{2023}]{jain2023helioseismic}
\begin{barticle}
\bauthor{\bsnm{Jain}, \binits{K.}},
\bauthor{\bsnm{Chowdhury}, \binits{P.}},
\bauthor{\bsnm{Tripathy}, \binits{S.C.}}:
\byear{2023},
\batitle{Helioseismic Investigation of Quasi-biennial Oscillation Source Regions}.
\bjtitle{The Astrophysical Journal}
\bvolume{959},
\bfpage{16}.
\end{barticle}
\endbibitem

\bibitem[\protect\citeauthoryear{Jefferies et~al.}{1988}]{jefferies1988helioseismology}
\begin{barticle}
\bauthor{\bsnm{Jefferies}, \binits{S.}},
\bauthor{\bsnm{Pomerantz}, \binits{M.}},
\bauthor{\bsnm{Duvall~Jr}, \binits{T.}},
\bauthor{\bsnm{Harvey}, \binits{J.}},
\bauthor{\bsnm{Jaksha}, \binits{D.}}:
\byear{1988},
\batitle{Helioseismology from the South Pole: comparison of 1987 and 1981 results.}
\bjtitle{Seismology of the Sun and Sun-like Stars}
\bvolume{286},
\bfpage{279}.
\end{barticle}
\endbibitem

\bibitem[\protect\citeauthoryear{Jim{\'e}nez}{2006}]{jimenez2006estimation}
\begin{barticle}
\bauthor{\bsnm{Jim{\'e}nez}, \binits{A.}}:
\byear{2006},
\batitle{An estimation of the acoustic cutoff frequency of the Sun based on the properties of the low-degree pseudomodes}.
\bjtitle{The Astrophysical Journal}
\bvolume{646},
\bfpage{1398}.
\end{barticle}
\endbibitem

\bibitem[\protect\citeauthoryear{Jim{\'e}nez-Reyes et~al.}{1998}]{jimenez1998solar}
\begin{barticle}
\bauthor{\bsnm{Jim{\'e}nez-Reyes}, \binits{S.}},
\bauthor{\bsnm{R{\'e}gulo}, \binits{C.}},
\bauthor{\bsnm{Pall{\'e}}, \binits{P.}},
\bauthor{\bsnm{Roca~Cort{\'e}s}, \binits{T.}}:
\byear{1998},
\batitle{Solar activity cycle frequency shifts of low-degree p-modes}.
\bjtitle{Astronomy and Astrophysics, v. 329, p. 1119-1124 (1998)}
\bvolume{329},
\bfpage{1119}.
\end{barticle}
\endbibitem

\bibitem[\protect\citeauthoryear{{Keith-Hardy}, {Tripathy}, and {Jain}}{2019}]{2019ApJ...877..148K}
\begin{barticle}
\bauthor{\bsnm{{Keith-Hardy}}, \binits{J.Z.}},
\bauthor{\bsnm{{Tripathy}}, \binits{S.C.}},
\bauthor{\bsnm{{Jain}}, \binits{K.}}:
\byear{2019},
\batitle{{Modeling the Effects of Observational Gaps on p-mode Oscillation Parameters}}.
\bjtitle{The Astrophysical Journal}
\bvolume{877},
\bfpage{148}.
\end{barticle}
\endbibitem

\bibitem[\protect\citeauthoryear{Kiefer et~al.}{2017}]{kiefer2017}
\begin{barticle}
\bauthor{\bsnm{Kiefer}, \binits{R.}},
\bauthor{\bsnm{Schad}, \binits{A.}},
\bauthor{\bsnm{Davies}, \binits{G.}},
\bauthor{\bsnm{Roth}, \binits{M.}}:
\byear{2017},
\batitle{Stellar magnetic activity and variability of oscillation parameters: An investigation of 24 solar-like stars observed by Kepler}.
\bjtitle{Astronomy and Astrophysics}
\bvolume{598},
\bfpage{A77}.
\end{barticle}
\endbibitem

\bibitem[\protect\citeauthoryear{{Kiefer} et~al.}{2018}]{2018SoPh..293..151K}
\begin{barticle}
\bauthor{\bsnm{{Kiefer}}, \binits{R.}},
\bauthor{\bsnm{{Komm}}, \binits{R.}},
\bauthor{\bsnm{{Hill}}, \binits{F.}},
\bauthor{\bsnm{{Broomhall}}, \binits{A.-M.}},
\bauthor{\bsnm{{Roth}}, \binits{M.}}:
\byear{2018},
\batitle{{GONG p-Mode Parameters Through Two Solar Cycles}}.
\bjtitle{Solar Physics}
\bvolume{293},
\bfpage{151}.
\end{barticle}
\endbibitem

\bibitem[\protect\citeauthoryear{Kosak, Kiefer, and Broomhall}{2022}]{kosak2022multi}
\begin{barticle}
\bauthor{\bsnm{Kosak}, \binits{K.}},
\bauthor{\bsnm{Kiefer}, \binits{R.}},
\bauthor{\bsnm{Broomhall}, \binits{A.-M.}}:
\byear{2022},
\batitle{A multi-instrument investigation of the frequency stability of oscillations above the acoustic cut-off frequency with solar activity}.
\bjtitle{Monthly Notices of the Royal Astronomical Society}
\bvolume{512},
\bfpage{5743}.
\end{barticle}
\endbibitem

\bibitem[\protect\citeauthoryear{Kumar and Lu}{1991}]{kumar1991location}
\begin{barticle}
\bauthor{\bsnm{Kumar}, \binits{P.}},
\bauthor{\bsnm{Lu}, \binits{E.}}:
\byear{1991},
\batitle{The location of the source of high-frequency solar acoustic oscillations}.
\bjtitle{The Astrophysical Journal}
\bvolume{375},
\bfpage{L35}.
\end{barticle}
\endbibitem

\bibitem[\protect\citeauthoryear{Kumar et~al.}{1990}]{kumar1990observed}
\begin{bchapter}
\bauthor{\bsnm{Kumar}, \binits{P.}},
\bauthor{\bsnm{Duvall}, \binits{T.}},
\bauthor{\bsnm{Harvey}, \binits{J.}},
\bauthor{\bsnm{Jefferies}, \binits{S.}},
\bauthor{\bsnm{Pomerantz}, \binits{M.}},
\bauthor{\bsnm{Thompson}, \binits{M.}}:
\byear{1990},
\bctitle{What are the observed high-frequency solar acoustic modes?}
In: \bbtitle{Progress of Seismology of the Sun and Stars: Proceedings of the Oji International Seminar Held at Hakone, Japan, 11--14 December 1989},
\bfpage{87}.
\bcomment{Springer}.
\end{bchapter}
\endbibitem

\bibitem[\protect\citeauthoryear{Libbrecht}{1988}]{libbrecht1988solar}
\begin{barticle}
\bauthor{\bsnm{Libbrecht}, \binits{K.}}:
\byear{1988},
\batitle{Solar p-mode phenomenology}.
\bjtitle{The Astrophysical Journal}
\bvolume{334},
\bfpage{510}.
\end{barticle}
\endbibitem

\bibitem[\protect\citeauthoryear{Libbrecht and Woodard}{1990}]{libbrecht1990solar}
\begin{barticle}
\bauthor{\bsnm{Libbrecht}, \binits{K.G.}},
\bauthor{\bsnm{Woodard}, \binits{M.}}:
\byear{1990},
\batitle{Solar-cycle effects on solar oscillation frequencies}.
\bjtitle{Nature}
\bvolume{345},
\bfpage{779}.
\end{barticle}
\endbibitem

\bibitem[\protect\citeauthoryear{Mehta et~al.}{2022}]{mehta2022cycle}
\begin{barticle}
\bauthor{\bsnm{Mehta}, \binits{T.}},
\bauthor{\bsnm{Jain}, \binits{K.}},
\bauthor{\bsnm{Tripathy}, \binits{S.}},
\bauthor{\bsnm{Kiefer}, \binits{R.}},
\bauthor{\bsnm{Kolotkov}, \binits{D.}},
\bauthor{\bsnm{Broomhall}, \binits{A.-M.}}:
\byear{2022},
\batitle{Cycle dependence of a quasi-biennial variability in the solar interior}.
\bjtitle{Monthly Notices of the Royal Astronomical Society}
\bvolume{515},
\bfpage{2415}.
\end{barticle}
\endbibitem

\bibitem[\protect\citeauthoryear{Nishizawa and Shibahashi}{1995}]{nishizawa1995implication}
\begin{bchapter}
\bauthor{\bsnm{Nishizawa}, \binits{Y.}},
\bauthor{\bsnm{Shibahashi}, \binits{H.}}:
\byear{1995},
\bctitle{Implication of long-term frequency variation}.
In: \bbtitle{GONG 1994. Helio-and Astro-Seismology from the Earth and Space}
\bseriesno{76},
\bfpage{280}.
\end{bchapter}
\endbibitem

\bibitem[\protect\citeauthoryear{Pall{\'e}, R{\'e}gulo, and Roca~Cort{\'e}s}{1989}]{palle1989solar}
\begin{barticle}
\bauthor{\bsnm{Pall{\'e}}, \binits{P.}},
\bauthor{\bsnm{R{\'e}gulo}, \binits{C.}},
\bauthor{\bsnm{Roca~Cort{\'e}s}, \binits{T.}}:
\byear{1989},
\batitle{Solar cycle induced variations of the low L solar acoustic spectrum}.
\bjtitle{Astronomy and Astrophysics}
\bvolume{224},
\bfpage{253}.
\end{barticle}
\endbibitem

\bibitem[\protect\citeauthoryear{Pall{\'e}, R{\'e}gulo, and Roca~Cort{\'e}s}{1990}]{palle1990progress}
\begin{barticle}
\bauthor{\bsnm{Pall{\'e}}, \binits{P.}},
\bauthor{\bsnm{R{\'e}gulo}, \binits{C.}},
\bauthor{\bsnm{Roca~Cort{\'e}s}, \binits{T.}}:
\byear{1990},
\batitle{Progress of Seismology of the Sun and Stars}.
\bjtitle{Lectures Notes in Physics}
\bvolume{367},
\bfpage{129}.
\end{barticle}
\endbibitem

\bibitem[\protect\citeauthoryear{Rhodes et~al.}{2011}]{rhodes2011temporal}
\begin{bchapter}
\bauthor{\bsnm{Rhodes}, \binits{E.}},
\bauthor{\bsnm{Reiter}, \binits{J.}},
\bauthor{\bsnm{Schou}, \binits{J.}},
\bauthor{\bsnm{Larson}, \binits{T.}},
\bauthor{\bsnm{Scherrer}, \binits{P.}},
\bauthor{\bsnm{Brooks}, \binits{J.}},
\bauthor{\bsnm{McFaddin}, \binits{P.}},
\bauthor{\bsnm{Miller}, \binits{B.}},
\bauthor{\bsnm{Rodriguez}, \binits{J.}},
\bauthor{\bsnm{Yoo}, \binits{J.}}:
\byear{2011},
\bctitle{Temporal changes in the frequencies and widths of the solar p-mode oscillations}.
In: \bbtitle{Journal of Physics: Conference Series}
\bseriesno{271}.
\end{bchapter}
\endbibitem

\bibitem[\protect\citeauthoryear{Ronan, Cadora, and Labonte}{1994}]{ronan1994solar}
\begin{barticle}
\bauthor{\bsnm{Ronan}, \binits{R.}},
\bauthor{\bsnm{Cadora}, \binits{K.}},
\bauthor{\bsnm{Labonte}, \binits{B.}}:
\byear{1994},
\batitle{Solar cycle changes in the high frequency spectrum}.
\bjtitle{Solar Physics}
\bvolume{150},
\bfpage{389}.
\end{barticle}
\endbibitem

\bibitem[\protect\citeauthoryear{Simoniello et~al.}{2012}]{simoniello2012seismic}
\begin{barticle}
\bauthor{\bsnm{Simoniello}, \binits{R.}},
\bauthor{\bsnm{Jain}, \binits{K.}},
\bauthor{\bsnm{Tripathy}, \binits{S.}},
\bauthor{\bsnm{Turck-Chi{\'e}ze}, \binits{S.}},
\bauthor{\bsnm{Finsterle}, \binits{W.}},
\bauthor{\bsnm{Roth}, \binits{M.}}:
\byear{2012},
\batitle{Seismic comparison of the 11-and 2-year cycle signatures in the Sun}.
\bjtitle{Astronomische Nachrichten}
\bvolume{333},
\bfpage{1018}.
\end{barticle}
\endbibitem

\bibitem[\protect\citeauthoryear{Simoniello et~al.}{2016}]{simoniello2016new}
\begin{barticle}
\bauthor{\bsnm{Simoniello}, \binits{R.}},
\bauthor{\bsnm{Tripathy}, \binits{S.}},
\bauthor{\bsnm{Jain}, \binits{K.}},
\bauthor{\bsnm{Hill}, \binits{F.}}:
\byear{2016},
\batitle{A new challenge to solar dynamo models from helioseismic observations: The latitudinal dependence of the progression of the solar cycle}.
\bjtitle{The Astrophysical Journal}
\bvolume{828},
\bfpage{41}.
\end{barticle}
\endbibitem

\bibitem[\protect\citeauthoryear{Tapping}{2013}]{tapping201310}
\begin{barticle}
\bauthor{\bsnm{Tapping}, \binits{K.}}:
\byear{2013},
\batitle{The 10.7 cm solar radio flux (F10. 7)}.
\bjtitle{Space Weather}
\bvolume{11},
\bfpage{394}.
\end{barticle}
\endbibitem

\bibitem[\protect\citeauthoryear{Virtanen}{2020}]{2020SciPy-NMeth}
\begin{barticle}
\bauthor{\bsnm{Virtanen}, \binits{P.}}:
\byear{2020},
\batitle{{{SciPy} 1.0: Fundamental Algorithms for Scientific Computing in Python}}.
\bjtitle{Nature Methods}
\bvolume{17},
\bfpage{261}.
\end{barticle}
\endbibitem

\bibitem[\protect\citeauthoryear{Vorontsov et~al.}{1998}]{vorontsov1998acoustic}
\begin{barticle}
\bauthor{\bsnm{Vorontsov}, \binits{S.}},
\bauthor{\bsnm{Jefferies}, \binits{S.}},
\bauthor{\bsnm{Duval~Jr}, \binits{T.}},
\bauthor{\bsnm{Harvey}, \binits{J.}}:
\byear{1998},
\batitle{Acoustic interferometry of the solar atmosphere: p-modes with frequencies near the ‘acoustic cut-off’}.
\bjtitle{Monthly Notices of the Royal Astronomical Society}
\bvolume{298},
\bfpage{464}.
\end{barticle}
\endbibitem

\bibitem[\protect\citeauthoryear{Woodard and Noyes}{1985}]{woodard1985change}
\begin{barticle}
\bauthor{\bsnm{Woodard}, \binits{M.F.}},
\bauthor{\bsnm{Noyes}, \binits{R.W.}}:
\byear{1985},
\batitle{Change of solar oscillation eigenfrequencies with the solar cycle}.
\bjtitle{Nature}
\bvolume{318},
\bfpage{449}.
\end{barticle}
\endbibitem

\end{thebibliography}

\end{document}